\documentclass[10pt,journal]{IEEEtran}

\IEEEoverridecommandlockouts

\usepackage{amsmath,amsfonts,amssymb,color}
\usepackage{amsthm}
\usepackage{epsfig}
\usepackage{psfrag}
\usepackage{algorithm}
\usepackage{algpseudocode}
\usepackage{array}
\usepackage{epstopdf}
\usepackage{tikz}
\usepackage{float}
\usepackage{inputenc}
\usepackage{tikz, subcaption, cite}
\usepackage{multicol,multirow}
\usepackage{algorithmicx}
\usepackage{algpseudocode}

\newtheorem{problem}{Problem}
\newtheorem{theorem}{Theorem}
\newtheorem{corollary}{Corollary}
\newtheorem{lemma}{Lemma}
\newtheorem{remark}{Remark}
\newtheorem{definition}{Definition}
\newtheorem{proposition}{Proposition}
\newtheorem{example}{Example}

\begin{document}
\title{Optimal Policies for Recovery of Multiple Systems After Disruptions}
\author{Hemant Gehlot, Shreyas Sundaram and Satish V. Ukkusuri
\thanks{Hemant Gehlot and Satish V. Ukkusuri are with the Lyles School of Civil Engineering at Purdue University. Email: {\tt \{hgehlot,sukkusur\}@purdue.edu}. Shreyas Sundaram is with the School of Electrical and Computer Engineering at Purdue University. Email: {\tt sundara2@purdue.edu}.  This research was supported by National Science Foundation award CMMI 1638311.}
}

\maketitle
\begin{abstract}
We consider a scenario where a system experiences a disruption, and the states (representing health values) of its components continue to reduce over time, unless they are acted upon by a controller. 
Given this dynamical setting, we consider the problem of finding an optimal control (or switching)  sequence to maximize the sum of the weights of the components whose states are brought back to the maximum value.
We first provide several characteristics of the optimal policy for the general (fully heterogeneous) version of this problem.  
We then show that under certain conditions on the rates of repair and deterioration, we can explicitly characterize the optimal control policy as a function of the states. 
When the deterioration rate (when not being repaired) is larger than or equal to the repair rate, and the deterioration and repair rates as well as the weights are homogeneous across all the components, the optimal control policy is to target the component that has the largest state value at each time step. On the other hand, if the repair rates are sufficiently larger than the deterioration rates, the optimal control policy is to target the component whose state minus the deterioration rate is least in a particular subset of components at each time step. 
\end{abstract}

\section{Introduction}
We study a control problem where a set of components of a system are damaged after a disruptive event (such as a natural disaster or security breach), and their health values (or states) continue to deteriorate over time unless they are repaired. In the absence of intervention, these components will eventually reach a {\it permanent failure} state. An entity (or controller) is responsible for counteracting the deterioration process by targeting the components for repair; this increases the states of the components to a value known as {\it permanent repair}. The state of each component does not change once it reaches either of the two thresholds (permanent repair or failure). Under these dynamics, the entity needs to make optimal control decisions for repairing different components to maximize a performance criterion or reward, e.g., maximizing the number or reward of components that are permanently repaired. This problem has applications in multiple areas including  post-disaster recovery, protection of cyber-physical systems against attacks, fire fighting, epidemic control, etc. For instance, infrastructure components face accelerated deterioration after disasters  due to processes such as floods and corrosion \cite{gaspard2007impact}, and can deteriorate to such a level that they become unusable and require full replacement, which is expensive and thus undesirable. Similarly, when multiple computer servers are infiltrated by an attacker or virus, the protecting agency has a limited amount of time before the servers become fully compromised \cite{leversage2008estimating}. Likewise, in forest fires, the objective of fire-fighters is to ensure that the fire does not enter a state known as \textit{flashover}, where there is little hope of saving the affected property or individuals \cite{metropolitan2009fire}. Therefore, shortage of available personnel and resources require the fire-fighters to make optimal decisions to control simultaneous fires located in different regions \cite{smeby2013fire}. 

Our problem falls into the general class of optimal control and scheduling of switched systems \cite{zhang2009value,gorges2010optimal} (or, more generally, hybrid systems \cite{borrelli2005dynamic,bemporad2006logic}). A switched system consists of multiple subsystems that are governed by different dynamical rules such that only one subsystem is active at each point of time. In our problem, each component corresponds to a subsystem, and the switching rule corresponds to which component the controller chooses to target for repair at each time-step.  Since the entity chooses which component to target for repair at fixed intervals of time (e.g., on an hourly or daily basis in the case of natural disasters), our problem comes under the class of discrete-time switched systems. The main source of complexity in the optimal control and scheduling of discrete-time switched systems is the combinatorial number of feasible switching sequences \cite{bemporad2006logic}. In the past decade, there have been some advances in theoretical results and computational frameworks for solving switched systems. For example, the papers \cite{zhang2009value,gorges2010optimal} characterize optimal/near-optimal control and scheduling policies for discrete-time switched linear systems with linear or quadratic cost/reward functions.  However, there are no theoretical results or computational frameworks that efficiently solve the optimal control and scheduling problems for all types of switched systems and most results are formulation dependent \cite{zhu2015optimal}. Therefore, optimal control and scheduling of switched systems remains an area of active research. 

\subsection*{Contributions of our paper}
For the setting described above, we consider the problem of finding the optimal switching policy to optimize a reward function given by the sum of the weights of components that are permanently repaired. 
We find that the optimal switching policies are state feedback policies that depend on the relationship between the rates of repair (when being targeted by the controller) and deterioration (when not being targeted) of the health values of the components. Specifically, the contributions of this paper are as follows.  First, when the deterioration rates are larger than or equal to the repair rates, we prove that the optimal switching policy is to permanently repair a component before switching to target another component. We also show that when the repair and deterioration rates satisfy a certain condition, the optimal policy can be found in polynomial time (although the exponent of the polynomial can be large). Second, when the repair and deterioration rates (and the weights) are homogeneous across all components, we explicitly characterize the optimal policy to be the one that targets the damaged component with the largest state (i.e., health) at each time-step. We also show that such a policy provides an approximately optimal solution when the component weights are heterogeneous.  
 Third, when the repair rates are sufficiently larger than the deterioration rates, we prove that the optimal switching policy is to target the component whose health minus the rate of deterioration is smallest in a particular subset of components at each time-step. 

\subsection*{Relationship to existing literature}
At a high level, other switched system control problems such as scheduling of thermostatically controlled loads \cite{nghiem2011green,nilsson2017class} also have similarities to our problem. These studies characterize scheduling control policies so that the states  (e.g., temperature) of the components (e.g., rooms) in the system always stay in a given interval.  
In these studies, the system becomes unstable (equivalent to the notion of permanent failure in our problem) if the state of a component violates any of the two interval thresholds. In contrast, our problem has one desirable threshold and one undesirable/failure threshold. This difference leads us to characterize optimal policies of different types depending on the problem conditions. For example, we show that non-jumping policies, where switching between different components is not allowed until a component is permanently repaired, turn out to be optimal under some conditions. In contrast, the aforementioned studies related to switched systems do not characterize non-jumping policies to be optimal; indeed, jumping is necessary to meet the objectives of those problems. 

Our problem is motivated by the fact that after shocks/disruptions, components such as infrastructure, infected servers, fire-affected regions etc., deteriorate rapidly in comparison to the deterioration faced during normal times \cite{gaspard2007impact,smeby2013fire}. Thus, it can be assumed that the state of a component does not significantly vary due to normal deterioration processes once the state of permanent repair is reached. Analogies to such a model can also be found in model predictive control problems where the objective of the problems is to ensure that the components' states lie within a set of desirable states and there is an associated penalty cost if the states reach an undesirable value  \cite{scokaert1999feasibility,mhaskar2006stabilization}. Similarly, the objective of our problem is to maximize the sum of the weights of the components whose states can be brought back to the desirable threshold value (permanent repair) without ever reaching the undesirable threshold (permanent failure). 

Problems of a similar flavor can also be found in optimal control of robotic systems \cite{smith2012persistent} that persistently monitor changing environments; there, the goal is to keep the level of uncertainty about some dynamic phenomenon below a certain threshold, with the uncertainty growing over time whenever the phenomenon is not being observed. Our problem also has similarities to the problem of allocating resources (e.g., time slots) at a base station to many time-varying competing flows/queues \cite{eryilmaz2007fair,eryilmaz2005stable,buche2004control}. However, these studies do not consider permanent failure of components or flows being serviced, and instead focus on either bounding the long-term state of the system, or maximizing long-term throughput or stability. Job scheduling problems with degrading processing times \cite{cheng2004concise,wei2012single,ng2010preemptive} as a function of job starting times also have analogies to our problem. A major difference between job scheduling and our problem is that in the former, a job is considered to be late if its \textit{completion} time exceeds the corresponding due date, whereas in our problem, a component is considered to be failed if its health reaches the state of permanent failure before the entity \textit{starts} to control it. Another important difference is that a job is completely processed even if its completion time exceeds the corresponding due date; in contrast, a component in our problem cannot be targeted if its health value reaches the state of permanent failure.
Our problem also has similarities with scheduling analysis of real-time systems \cite{zhang2009schedulability}; there, the analysis focuses on real-time tasks that become available for processing at different times; in contrast, all the components in our problem are available for control starting at the same time. Patient triage scheduling problems \cite{argon2008scheduling} also have some analogies to our problem; these problems only focus on non-jumping sequences and characterize optimal sequences assuming that if a task has less time left before expiration than another task, then the former task also takes less time to be completed than the latter task. However, this assumption does not hold for the problem that we consider in this paper. 

In the conference version of this paper \cite{gehlot2019optimal}, we considered the setting where the weights and the rates of repair and deterioration are homogeneous across all the components. This paper significantly expands upon the conference paper by considering heterogeneous rates and weights, and showing that it is not optimal to switch away from a component before permanently repairing it when the deterioration rates are larger than the repair rates (even for heterogeneous rates and weights across the components), and fully characterizing the optimal control policy when the rates of repair are sufficiently larger than the rates of deterioration (even for heterogeneous rates and weights across the components).  
The outline of the paper is as follows. In the next section, we formally present the problem that we consider in this paper. After this, we 
characterize the optimal control policies for certain instances of this problem. Finally, we present the results of simulation studies to compare the optimal control policies with randomly generated control policies.     

\section{Problem Statement}
\label{probdesc}
There are $N (\geq 2)$ nodes indexed by the set $\mathcal{V} = \{1, 2, \ldots, N\}$, each representing a component (depending on the context, this could be a portion of physical infrastructure in a given area, an infected computer server, a fire-affected region etc.).  
There is an entity (or controller) whose objective is to repair these components. We assume that time progresses in discrete time steps, capturing the resolution at which the entity makes decisions about which node to repair. We index the time steps with the variable $t \in \mathbb{N} = \{0, 1, 2, \ldots\}$. 
The {\it health} of each node $j \in \mathcal{V}$ at time step $t$ is denoted by $v^j_t \in [0,1]$.\footnote{As mentioned earlier, the health values/states of the nodes are bounded by two thresholds: \textit{permanent repair} and \textit{permanent failure}. Therefore, by scaling the health values and repair/deterioration rates, we can take the range of health values to be the interval [0,1] without loss of generality.} The initial health of each node $j$ is denoted by $v_0^j \in (0,1)$. The aggregate state vector for the entire system at each time step $t \in \mathbb{N}$ is given by $\{v^j_t\}$, where $j\in\{1,\ldots,N\}$. The weight of node $j\in\{1,\ldots,N\}$ is denoted by $w^j \in \mathbb R_{\ge 0}$,  and represents its relative importance.  For example, it can represent the number of households that are dependent on an infrastructure component, the number of files that are stored in a computer server, or the population of a fire-affected region.  

\begin{definition}
We say that node $j$ \textbf{permanently fails} at time step $t$ if $v^j_t = 0$ and $v^j_{t-1} > 0$. We say that node $j$ is \textbf{permanently repaired} at time step $t$ if $v^j_t = 1$ and $v^j_{t-1} < 1$. If a node permanently fails or is permanently repaired, then its health does not change thereafter.
\end{definition}

At each time step $t$, the entity can target exactly one node to repair during that time step.\footnote{We leave an investigation of the case where the entity can simultaneously target multiple nodes for future work.} Thus, the control action taken by the entity at time step $t$ is denoted by $u_t \in \mathcal{V}$. If node $j$ is being repaired by the entity at time step $t$ and it has not permanently failed or repaired, its health increases by a quantity $\Delta_{inc}^j \in [0,1]$ (up to a maximum health of $1$). If node $j$ is not being repaired by the entity at time step $t$ and it has not permanently failed or repaired, its health decreases by a fixed quantity\footnote{We make the assumption of constant repair and deterioration rates for analytical tractability. However, we anticipate that some of the results presented in this paper can be extended to non-constant rates; we keep the exploration of this case as a future extension.} $\Delta_{dec}^j \in [0,1]$ (down to a minimum health of $0$). Thus, $\{\Delta_{inc}^j\}$ and $\{\Delta_{dec}^j\}$ represent the vectors of the rates of repair and deterioration, respectively. For each node $j$, the dynamics of the control problem are given by
\begin{equation}
v^j_{t+1} = \begin{cases} 1 &\text{if } v^j_t=1, \\
0 &\text{if } v^j_t=0, \\
\min(1,v^j_t + \Delta_{inc}^j) & \hbox{if } u_t = j \hspace{1mm}\text{and}\hspace{1mm} v^j_t\in(0,1), \\
\max(0, v^j_t-\Delta_{dec}^j) & \hbox{if } u_t \neq j \hspace{1mm}\text{and}\hspace{1mm} v^j_t\in(0,1).\end{cases}
\label{eq:controlled_dynamics}
\end{equation}

\begin{definition}
For any given initial state values $v_0 = \{v_0^j\}$, weights $w=\{w^j\}$, and control sequence $U = \{u_{0},u_{1},\ldots\}$, let $\mathcal{M}(v_0, U)$ be the set of nodes that are permanently repaired through that sequence. That is, $\mathcal{M}(v_0, U)= \{j\in \mathcal{V}\hspace{1mm}|\hspace{1mm}\exists \hspace{1mm}t\ge 0 \text{ s.t. }  v^j_t=1\}$. We define the \textbf{reward} $J(v_0, w,U)$ as the sum of the weights of the nodes in set $\mathcal{M}(v_0,U)$. Mathematically, $J(v_0,w, U)=\sum_{j\in \mathcal{M}(v_0, U)} w^j$. 
\label{def:reward}
\end{definition}

Based on the dynamics \eqref{eq:controlled_dynamics} and the reward definition given above, we study the following problem in this paper.

\begin{problem} \label{problem}
Given a set $\mathcal{V}$ of $N$ nodes with initial health values $v_0 = \{v_0^j\}$, weights $w=\{w^j\}$, repair rates $\{\Delta_{inc}^j\}$, and deterioration rates $\{\Delta_{dec}^j\}$, find a control sequence $U = \{u_0, u_1, \ldots\}$ that maximizes the reward $J(v_0, w,U)$.
\end{problem}

Before presenting our analysis of the problem, we introduce the concept of a \textbf{jump}.  
\begin{definition}
The entity is said to have jumped at some time step $t$ if it starts targeting a different node before permanently repairing the node it targeted in the last time step. That is, if $u_{t-1}=j$, $v_{t}^j<1$ and $u_{t} \neq j$ then the entity is said to have jumped at time step $t$. A control sequence that does not contain any jumps is said to be a \textbf{non-jumping sequence}. 
\end{definition}

We will  split our analysis of the optimal control policy for Problem~\ref{problem} into two parts: one for the case where $\Delta_{dec}^j \geq \Delta_{inc}^j, \forall j\in\{1,\ldots,N\}$, and the other for the remaining cases. 

\section{Optimal Control Policies for $\Delta_{dec}^j \geq \Delta_{inc}^j,$ $ \forall j \in \{1,\ldots,N\}$}
We first show that non-jumping policies are optimal when $\Delta_{dec}^j\ge \Delta_{inc}^j, \forall j\in \{1,\ldots,N\}$. 
Subsequently, we show that when the repair rates are lower bounded by a positive real number, the optimal control sequence can be found via an algorithm that has run-time polynomial in the number of nodes (but exponential in a certain function of the repair and deterioration rates). After this, we consider the special case of Problem \ref{problem} where the weights and rates are homogeneous across all the nodes. For this case, we explicitly characterize the optimal non-jumping policy, and thereby find the globally optimal policy. Finally, we characterize an efficient algorithm to compute an approximately optimal solution when the rates are homogeneous but the weights are heterogeneous.  

\subsection{Optimality of non-jumping policies}
First, we analyze properties of sequences containing at most one jump and later generalize to sequences containing an arbitrary number of jumps. We start with the following result.

\begin{figure}[H]
	\begin{center}
		\includegraphics[scale=0.6]{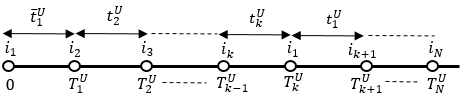}
	\end{center}	
	\caption{Sequence $U$ with a single jump.}
	\label{seqU}
\end{figure}

\begin{figure}[H]
	\begin{center}
		\includegraphics[scale=0.6]{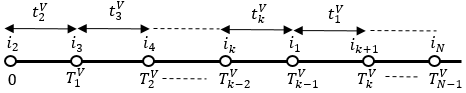} 
	\end{center}	
	\caption{Non-jumping sequence $V$.}
	\label{seqV}
\end{figure}

\begin{lemma} \label{timecomparison}
Let there be $N (\geq 2)$ nodes, and suppose $\Delta_{dec}^j\ge\Delta_{inc}^j, \forall j\in\{1,\ldots,N\}$. Consider the two control sequences $U$ and $V$ targeting $N$ nodes shown in Figures \ref{seqU} and \ref{seqV}, respectively. Suppose sequence $U$ permanently repairs all nodes and contains exactly one jump, where the entity partially repairs node $i_1$ before moving to node $i_2$ at time step $\bar{t}_{1}^U$. Sequence $U$ then permanently repairs nodes $i_2, i_3, \ldots, i_k$, before returning to node $i_1$ and permanently repairing it. Sequence $V$ is a non-jumping sequence that targets nodes in the order $i_2, i_3, \ldots, i_k, i_1, i_{k+1}, \ldots, i_N$.  Let $t_j^U$ (resp. $t_j^V$) be the number of time steps taken to permanently repair node $i_j$ in sequence $U$ (resp. sequence $V$). Then, sequence $V$ also permanently repairs all nodes, and furthermore, the following holds true:
\begin{align}
t_j^U &\geq t_j^V+\left(2^{j-2}\right)\overline{t}_1^U \hspace{3mm}\forall j \in \{2,\ldots,k\}, \label{claim1} \\
t_1^U &\geq t_1^V+(2^{k-1}-2)\overline{t}_1^U, \label{claim2} \\
t_j^U &\geq t_j^V+\left(2^{j-1}-2^{j-k}\right)\overline{t}_1^U \hspace{3mm}\forall j \in \{k+1,\ldots,N\}. \label{claim3}
\end{align}
\end{lemma}


\begin{IEEEproof}
Let $T_1^U, T_2^U, \ldots, T_N^U$ be the time steps at which sequence $U$ starts targeting a new node, as shown in Fig.~\ref{seqU}.  Similarly, let $T_1^V, T_2^V, \ldots, T_{N-1}^V$ be the time steps at which sequence $V$ starts targeting a new node, as shown in Fig.~\ref{seqV}.

We start by first proving condition \eqref{claim1}, using mathematical induction on the index of nodes in the sequence. Consider $j=2$. At time step $T_1^U$ in sequence $U$, the health of node $i_2$ is given by
\begin{equation*}
v^{i_2}_{T_1^U}=v^{i_2}_0-\Delta_{dec}^{i_2}\overline{t}_1^U.
\end{equation*}   
We now calculate $t_2^U$ as
\begin{align*}
t_2^U=\left\lceil \frac{1-v^{i_2}_{T_1^U}}{\Delta_{inc}^{i_2}}\right \rceil &=  \left\lceil \frac{1-v^{i_2}_0+\Delta_{dec}^{i_2}\overline{t}_1^U}{\Delta_{inc}^{i_2}}\right \rceil \\
&\geq \left\lceil \frac{1-v^{i_2}_{0}}{\Delta_{inc}^{i_2}}\right \rceil+\overline{t}_1^U =t_2^V+\overline{t}_1^U,
\end{align*}
which satisfies condition \eqref{claim1}. Suppose that condition \eqref{claim1} holds for $r$ nodes where $r<k$.  If sequence $U$ permanently repairs nodes $i_2, \ldots, i_r$, then so  does sequence $V$ (as each node is reached at an earlier time step in sequence $V$ than in sequence $U$, by the above inductive assumption).  We now compute $v^{i_{r+1}}_{T_r^U}$:
\begin{align*}
v^{i_{r+1}}_{T_r^U} &= v^{i_{r+1}}_{0}-\Delta_{dec}^{i_{r+1}}T_r^U \\
&=  v^{i_{r+1}}_{0}-\Delta_{dec}^{i_{r+1}}\left(\overline{t}_1^U+t_2^U+\ldots+t_r^U\right).
\end{align*}
Thus,
\begin{align*}
t_{r+1}^U &= \left\lceil \frac{1-v^{i_{r+1}}_{0}+\Delta_{dec}^{i_{r+1}}\left(\overline{t}_1^U+t_2^U+\ldots+t_r^U\right)}{\Delta_{inc}^{i_{r+1}}} \right\rceil\\
&\geq \left \lceil \frac{1-v^{i_{r+1}}_{0}+\Delta_{dec}^{i_{r+1}}\left(t_2^V+\ldots+t_r^V\right)}{\Delta_{inc}^{i_{r+1}}} +\right. \\ 
& \qquad \left. \frac{\Delta_{dec}^{i_{r+1}}\left(\overline{t}_1^U+\overline{t}_1^U+2\overline{t}_1^U+\ldots +2^{r-2}\overline{t}_1^U\right)}{\Delta_{inc}^{i_{r+1}}}\right\rceil \\
&\geq t_{r+1}^V+2^{r-1}\overline{t}_1^U.
\end{align*}
So, we have shown condition \eqref{claim1} by induction. We now prove condition \eqref{claim2}. Node $i_1$ is targeted again in sequence $U$ at time step $T_k^U$, at which point its health is
\begin{equation}
v^{i_1}_{T_k^U}=v^{i_1}_{0}+\overline{t}_1^U\Delta_{inc}^{i_1}-\left(t_2^U+\ldots+t_k^U\right)\Delta_{dec}^{i_1}.
\label{node1healthseqU}
\end{equation}
Thus, the number of time steps taken to permanently repair node $i_1$ in sequence $U$ (the second time it is targeted in the sequence) is 
\begin{equation}
t_1^U=\left\lceil \frac{1-v^{i_1}_0-\overline{t}_1^U\Delta_{inc}^{i_1}+\left(t_2^U+\ldots+t_k^U\right)\Delta_{dec}^{i_1}}{\Delta_{inc}^{i_1}}\right \rceil. 
\label{t1U}
\end{equation}
Note that
\begin{equation}
t_2^U+\ldots+t_k^U \geq 
t_2^V+\ldots+t_k^V+\left(2^{k-1}-1\right)\overline{t}_1^U, \label{t2Ulowerbound}
\end{equation}
by condition \eqref{claim1}. Furthermore, in sequence $V$, the health of node $i_1$ at the time it is targeted is given by
$$
v^{i_1}_{T_{k-1}^V} = v^{i_1}_0 - \Delta_{dec}^{i_1}(t_2^V + t_3^V + \cdots + t_k^V).
$$
Comparing this to the health of node $i_1$ in sequence $U$ at the time it is targeted (given by \eqref{node1healthseqU}), and using \eqref{t2Ulowerbound}, we see that since $i_1$ is assumed to not have failed in sequence $U$, it will not have failed in sequence $V$ as well.  Thus, the number of time steps required to permanently repair $i_1$ in sequence $V$ is given by
\begin{equation}
t_1^V = \left\lceil \frac{1-v^{i_1}_0 + \Delta_{dec}^{i_1}(t_2^V + t_3^V + \cdots + t_k^V)}{\Delta_{inc}^{i_1}}\right\rceil.
\label{t1V}
\end{equation}
Thus, using \eqref{t1U}, \eqref{t2Ulowerbound} and \eqref{t1V}, we have 
\begin{equation*}
t_1^U\geq t_1^V+\left(2^{k-1}-2\right)\overline{t}_1^U,
\end{equation*}
proving condition \eqref{claim2}.

We now prove condition \eqref{claim3} via mathematical induction. Consider node $i_{k+1}$. At the time step when this node is targeted in sequence $U$, its health is
\begin{align*}
v^{i_{k+1}}_{T^U_{k+1}} &=v^{i_{k+1}}_{0}-\Delta_{dec}^{i_{k+1}}T^U_{k+1}\\
&=v^{i_{k+1}}_{0}-\Delta_{dec}^{i_{k+1}}\left(\overline{t}_1^U+t_2^U+\ldots+t_k^U+t_1^U\right).
\end{align*}
If node $i_{k+1}$ has not failed at this point in sequence $U$, it has also not failed when it is reached in sequence $V$ (as all nodes prior to $i_{k+1}$ are permanently repaired faster in sequence $V$ than in sequence $U$, as shown above).  Thus, using \eqref{claim1} and \eqref{claim2},
\begin{multline*}
t_{k+1}^U=\left \lceil \frac{1-v^{i_{k+1}}_{0}+\Delta_{dec}^{i_{k+1}}\left(\overline{t}_1^U+t_2^U+\ldots+t_k^U+t_1^U\right)}{\Delta_{inc}^{i_{k+1}}} \right \rceil \\ \geq \left \lceil \frac{1-v^{i_{k+1}}_{0}+\Delta_{dec}^{i_{k+1}}\left(t_2^V+\ldots+t_k^V+t_1^V\right)}{\Delta_{inc}^{i_{k+1}}} \right. \\ +\left. \frac{\Delta_{dec}^{i_{k+1}}\left(\overline{t}_1^U+\overline{t}_1^U+2\overline{t}_1^U+\ldots+2^{k-2}\overline{t}_1^U+\left(2^{k-1}-2\right)\overline{t}_1^U\right)}{\Delta_{inc}^{i_{k+1}}}\right \rceil \\ \geq t_{k+1}^V+\left(2^k-2\right)\overline{t}_1^U,
\end{multline*}
which satisfies condition \eqref{claim3}. Suppose  condition \eqref{claim3} holds for $j \in \{k+1,\ldots,r\},$ where $ r<N$. Consider node $i_{r+1}$. 
Then, a similar inductive argument can be used to show that
\begin{equation*}
t_{r+1}^U
\geq t_{r+1}^V+\left(2^r-2^{r+1-k}\right)\overline{t}_1^U.
\end{equation*}
This proves the third claim.
\end{IEEEproof}

The above result considered sequences containing exactly one jump. This leads us to the following key result pertaining to the optimal control policy when $\Delta_{dec}^j \ge \Delta_{inc}^j, \forall j\in\{1,\ldots,N\}$.

\begin{theorem} \label{thm:nonjumping_optimal_notallfixed}
Let there be $N (\geq 2)$ nodes, and suppose $\Delta_{dec}^j\geq \Delta_{inc}^j, \forall j\in\{1,\ldots,N\}$. If there is a sequence $U$ with one or more jumps that permanently repairs all the nodes of a set $\mathcal{Z} \subseteq \mathcal{V}$, 
then there exists a non-jumping sequence that permanently repairs all the nodes in set $\mathcal{Z}$ in less time. Thus, non-jumping sequences are optimal when $\Delta_{dec}^j \ge \Delta_{inc}^j, \forall j\in\{1,\ldots,N\}$.
\end{theorem}
\begin{IEEEproof}
We prove that given a sequence with an arbitrary number of jumps that permanently repairs a set $\mathcal{Z} \subseteq \mathcal{V}$, one can come up with a sequence that permanently repairs all the nodes in $\mathcal{Z}$, but has at least one fewer jump than the given sequence (and permanently repairs in less time than the given sequence). One can iteratively apply this result on the obtained sequences to eventually yield a non-jumping sequence that permanently repairs all the nodes in $\mathcal{Z}$ in less time as compared to the given sequence; thus, the reward obtained by the non-jumping sequence will be equal to the reward obtained by the given sequence consisting of an arbitrary number of jumps.

Consider the given sequence $U$ that permanently repairs a set $\mathcal{Z}$ of nodes and suppose $U$ contains one or more jumps. 
Remove all the nodes targeted by $U$ that are not permanently repaired. This gives a new sequence $V$ that only targets nodes in the set $\mathcal{Z}$. If $V$ does not contain any jumps, then we are done.  Otherwise, consider the {\it last} jump in $V$, and suppose it occurs at time step $T$. Denote the portion of the sequence $V$ from time step $T-1$ onwards by $W$, and denote the portion of the sequence $V$ from time step $0$ to time step $T-2$ by $W'$. Now, note that sequence $W$ contains exactly one jump. Thus, by Lemma \ref{timecomparison}, we can replace sequence $W$ with another sequence $X$ that contains no jumps and permanently repairs all nodes that are permanently repaired in $W$ in less time. Create a new sequence $V'$ by concatenating the sequence $W'$ and the sequence $X$. Thus, $V'$ is a sequence with one fewer jump than $V$,  and permanently repairs all the nodes in set $\mathcal{Z}$ and does so in less time. The first part of the result thus follows. The fact that non-jumping policies are optimal is then immediately obtained by considering $U$ to be any optimal policy.
\end{IEEEproof}

\subsection{Optimal sequencing when the repair rates are lower bounded by a positive real number}
We now show that the optimal sequence can be found in polynomial time under certain conditions on the repair and deterioration rates. We start with the following result.
\begin{lemma}\label{lem:maxnodes}
    Let there be $N(\ge2)$ nodes, and suppose $\Delta_{dec}^j\geq \Delta_{inc}^j, \forall j\in\{1,\ldots,N\}$. Define $n= \min_j \left \lfloor\frac{\Delta_{dec}^j}{\Delta_{inc}^j} \right \rfloor$. Then, the number of nodes that can be permanently repaired by a non-jumping sequence is upper bounded by  
    \begin{equation}
        L = \min\Bigg\{N,\left \lfloor \log_{(1+n)} \left( \frac{n}{\min_j \{\Delta_{dec}^j \}} +1\right)+1 \right \rfloor \Bigg\}.
        \label{eq:max_nodes_fixed}
    \end{equation}
\end{lemma}

\begin{IEEEproof}
Theorem \ref{thm:nonjumping_optimal_notallfixed} showed that non-jumping sequences are optimal when $\Delta_{dec}^j\geq \Delta_{inc}^j, \forall j$.  
Next, note from the definition of $n$ that for each time step that a node $j$ deteriorates (where its health decreases by $\Delta_{dec}^j$), it will take at least $n$ time steps of repair to  compensate for that deterioration. We can now bound the number of nodes that are permanently repaired by a non-jumping sequence as follows. The number of time steps taken to permanently repair the first node is at least equal to $1$.  Then, the number of time steps taken to permanently repair the second node in the sequence is at least equal to $1+n$ (for the second node in the sequence, it takes at least $n$ time steps to repair the health that is lost due to deterioration and it takes at least one additional time step to repair the difference between the initial health and the permanent repair state). The number of time steps taken to permanently repair the third node in the sequence is at least equal to $1+n(1+1+n)$, i.e., $n$ times the number of time steps spent on repairing the previous nodes in order to make up for the deterioration faced in those time steps, and at least one additional time step to permanently repair. By induction, it can be easily shown that the number of time steps taken to permanently repair the $i_j$th node in the sequence is at least equal to $(1+n)^{j-1}$.
Suppose there exists a non-jumping sequence that permanently repairs $x$ nodes. Then, node $i_x$ in that sequence should have positive health by the time the first $x-1$ nodes are permanently repaired. The largest time step at which there is a node with positive health is upper bounded by $\max_j \bigg\{ \frac{v_0^j}{\Delta_{dec}^j} \bigg\}$. Then, $(1+n)^0+(1+n)^1+\ldots+(1+n)^{x-2}=\frac{(1+n)^{x-1}-1}{n} < \max_j \bigg\{ \frac{v_0^j}{\Delta_{dec}^j} \bigg\}$. Note that $\max_j \bigg\{ \frac{v_0^j}{\Delta_{dec}^j} \bigg\}<\max_j \bigg\{ \frac{1}{\Delta_{dec}^j} \bigg\}= \frac{1}{ \min_j \{ \Delta_{dec}^j \}} $ because $v_0^j<1, \forall j$. Thus, $x<\log_{(1+n)} \left( \frac{n}{\min_j \{\Delta_{dec}^j \}} +1\right)+1 $. 
\end{IEEEproof}

We now show that if $L$ in $\eqref{eq:max_nodes_fixed}$ is upper bounded (which will happen when the ratio $\frac{n}{\min_j \{\Delta_{dec}^j \}}$ is upper bounded by a positive real number), the optimal sequencing policy can be computed in time that is polynomial in the number of nodes. 

\begin{theorem} \label{thm:largenodes}
    Let there be $N(\ge2)$ nodes, and suppose $\Delta_{dec}^j\geq \Delta_{inc}^j, \forall j\in\{1,\ldots,N\}$. Define $n= \min_j \left \lfloor\frac{\Delta_{dec}^j}{\Delta_{inc}^j} \right \rfloor$. If $\frac{n}{\min_j \{\Delta_{dec}^j \}}$ is upper bounded by a positive real number, then the complexity of finding the optimal sequence is polynomial in the number of nodes. 
\end{theorem}

\begin{IEEEproof}
Since $\frac{n}{\min_j \{\Delta_{dec}^j \}}$ is upper bounded by a positive real number, the quantity $L$ in \eqref{eq:max_nodes_fixed} is upper-bounded by a constant.   
Under this condition, we enumerate all the non-jumping sequences of length $L$ that need to be compared to find the optimal sequence. At the start of the first time step, there are $N$ choices of nodes that can be targeted; after permanently repairing the first node, there are $N-1$ choices of nodes that can be permanently repaired, and so on. Since the maximum number of nodes that can be permanently repaired is upper bounded by $L$, the number of sequences that need to be compared to find the optimal sequence is $O\left(N^L\right)$. Denote the set of non-jumping sequences of length $L$ by  $\mathcal{W}$. We compute the weighted number of nodes that are permanently repaired by the sequences in set $\mathcal{W}$ through simulation. Since a sequence can permanently repair at most $L$ nodes, there would be $O(L)$ operations in the simulation while computing the weighted number of permanently repaired nodes. Thus, the complexity of computing the optimal sequence is $O\left(LN^L\right)$. Therefore, the complexity of finding the optimal sequence is polynomial in the number of nodes.
\end{IEEEproof}

\begin{remark} \label{rem:suffcient_mininc}
Note that the notation $O(LN^L)$ indicates that the complexity of computing the optimal sequence is upper bounded by a constant factor of $LN^L$ for {\it sufficiently large} $N$.  In particular, from \eqref{eq:max_nodes_fixed}, for sufficiently large $N$, $L$ will be given by the second term in the argument of the minimum function (under the condition in Theorem~\ref{thm:largenodes}), and thus will not grow with $N$.
In addition, note that $\frac{n}{\min_j \{\Delta_{dec}^j \}}=\frac{n}{\min_j \{n_j\Delta_{inc}^j \}}$, where $n_j\triangleq\frac{\Delta_{dec}^j}{\Delta_{inc}^j}\ge 1$. By definition, $n_j\ge n, \forall j $. Thus, $\frac{n}{\min_j \{n_j\Delta_{inc}^j \}}\le \frac{n}{\min_j \{n\Delta_{inc}^j \}}=\frac{1}{\min_j \{\Delta_{inc}^j \}} $. Therefore, a sufficient condition for $\frac{n}{\min_j \{\Delta_{dec}^j \}}$ to be upper bounded by a positive real number is that $\min_j \{\Delta_{inc}^j \}$ be lower bounded by a positive real number. Thus, the complexity of finding the optimal sequence is polynomial in the number of nodes if the repair rates are lower bounded by a positive real number and for all $j \in\{1,\ldots,N\}$, $\Delta_{dec}^j\geq \Delta_{inc}^j$. 
\end{remark}

While Theorem \ref{thm:largenodes} and Remark \ref{rem:suffcient_mininc} establish that the optimal sequence can be found in polynomial-time (specifically, $O(LN^L)$) if the repair rates are bounded away from zero and $\Delta_{dec}^j \ge \Delta_{inc}^j$ $\forall j \in \{1, 2, \ldots, N\}$, the exponent $L$ can be large if the repair rates are small.  In the next section, we focus on instances of Problem \ref{problem} where the weights and the rates of repair and deterioration are homogeneous. For such instances of the problem, we show that the optimal policy can be explicitly  characterized, regardless of the bound on the repair rates.   

\subsection{An optimal policy for homogeneous rates and weights}
We now consider a special case of Problem \ref{problem} when the deterioration and repair rates as well as the weights are homogeneous across all the nodes, i.e., $\Delta_{dec}^j=\Delta_{dec}, \forall j$, $\Delta_{inc}^j=\Delta_{inc}, \forall j$, and $w^j=\overline{w}, \forall j$. Theorem \ref{thm:nonjumping_optimal_notallfixed} showed that non-jumping policies are optimal for general (heterogeneous) rates and weights, and when the rates of deterioration are larger than the rates of repair, and thus this result holds for homogeneous rates and weights as well. For the homogeneous case, we characterize the optimal non-jumping policy in the set of all non-jumping policies. The following lemma will be useful for a later result.

\begin{lemma} \label{lemma1new}
Let there be $N (\geq2) $ nodes, and suppose for all $j\in\{1,\ldots,N\}$, $\Delta_{dec}^j=\Delta_{dec}$ and $\Delta_{inc}^j=\Delta_{inc}$. Consider a non-jumping sequence that permanently repairs all of the nodes. Under that sequence, suppose the order in which the nodes are targeted is $i_1,\ldots,i_N$ and that $t_{ j}$ is the number of time steps the entity takes to permanently repair node $i_j$.  Define $A_1=v_0^{i_1}$ and $A_k=v_0^{i_k} - \Delta_{dec} \sum_{j = 2}^{k} \left\lceil \frac{1-A_{j-1}}{\Delta_{inc}}\right\rceil$ for $ k \in \{2, \ldots, N\}$. Then, the following holds true:  
\begin{equation}
\sum_{p=1}^{N-1}t_p = \sum_{j = 2}^{N} \left\lceil \frac{1-A_{j-1}}{\Delta_{inc}}\right\rceil. \label{lemmaeq}
\end{equation}
\end{lemma}
The proof follows immediately from mathematical induction by noting that $A_j$ is the health of node $i_j$ when it is reached in the sequence, and thus $t_j=\left\lceil\frac{1-A_j}{\Delta_{inc}} \right\rceil, \forall j \in \{1,\ldots,N\}$.  

The next result presents the necessary and sufficient conditions for a non-jumping sequence to permanently repair all nodes.  

\begin{corollary} \label{necsuff_nowithin}
Let there be $N (\geq2) $ nodes, and suppose for all $j\in\{1,\ldots,N\}$, $\Delta_{inc}^j=\Delta_{inc}$ and $\Delta_{dec}^j=\Delta_{dec} $. Consider a non-jumping sequence, where the order in which the nodes are targeted is $i_1,\ldots,i_N$. Define $A_1=v_0^{i_1}$ and $A_k=v_0^{i_k} - \Delta_{dec} \sum_{j = 2}^{k} \left\lceil \frac{1-A_{j-1}}{\Delta_{inc}}\right\rceil$ for $k \in \{2,\ldots, N\}$. Then the following conditions are necessary and sufficient for all the nodes to eventually get permanently repaired:
\begin{equation}
A_k > 0 \hspace{3mm}\forall k \in \{1,\ldots, N\}. \label{ceil2}
\end{equation}
\end{corollary} 
The proof follows trivially from the definition of  $A_k$, namely that $A_k$ is the health of node $i_k$ at the time step when all nodes before $i_k$ in the sequence under consideration are permanently repaired and the entity starts repairing node $i_k$. 
	
Based on Corollary \ref{necsuff_nowithin}, we now provide the optimal policy  that permanently repairs the maximum number of nodes, under certain conditions on the initial health values, and rates of repair and deterioration.\footnote{Note that when weights are homogeneous across all the nodes,  Problem~\ref{problem} is equivalent to maximizing the number of nodes that are permanently repaired.}

\begin{theorem} \label{thm:multiplicityconditions_notfixall}
Let there be $N(\geq2)$ nodes, and suppose for all $j\in\{1,\ldots,N\}$, $\Delta_{inc}^j=\Delta_{inc}$, $\Delta_{dec}^j=\Delta_{dec}$, $w^j=\overline{w}$ and $\Delta_{dec}\ge \Delta_{inc}$. Suppose $\Delta_{dec} = n\Delta_{inc}$, where $n$ is a positive integer. Also, for each node $j\in\{1,\ldots,N\}$, suppose there exists a positive integer $m_j$ such that $1-v^{j}_0 =m_j \Delta_{inc}$. Then, the non-jumping sequence that targets nodes in decreasing order of their initial health is optimal for Problem~\ref{problem}. 
\end{theorem} 

\begin{IEEEproof}
Consider any optimal (non-jumping) sequence $U$, and let $x$ be the number of nodes that are permanently repaired by that sequence. Denote this set of $x(\le N)$ nodes as $\mathcal{Z}$. Let $i_1,\ldots,i_x$ be the order in which the sequence $U$ permanently repairs the $x$ nodes. The conditions $\Delta_{dec} = n\Delta_{inc}$ and $1-v^{j}_0 =m_j \Delta_{inc},\hspace{2mm} \forall j \in \{1,\ldots,N\}$ ensure that no node gets permanently repaired partway through a time step. Thus, the necessary and sufficient conditions to permanently repair $x$ nodes if a non-jumping sequence $U$ targets the nodes in the order $i_1,\ldots,i_x$ are given by $A_k>0, \forall k \in \{1,\ldots,x\}$ from Corollary \ref{necsuff_nowithin},
where $A_1=v^{i_1}_0$ and $A_k=v_0^{i_k} - n \sum_{j = 2}^{k} \left( 1-A_{j-1}\right)$ for $k\in \{2,\ldots,x\}$. Note that the ceiling functions in the definition of $A_k$ in Corollary~\ref{necsuff_nowithin} are dropped due to the conditions on the health values and the rates of repair and deterioration.  

As $\Delta_{dec}=n\Delta_{inc}$, we can expand these conditions as 
\begin{equation}
v_0^{i_k} - n\sum_{j = 2}^{k} \left((1-v_0^{i_{j-1}})(1+n)^{k-j}\right)>0 \hspace{3mm}\forall k \in \{2,\ldots, x\}. \label{iterative}
\end{equation}
The conditions \eqref{iterative} can be alternatively written as
\begin{equation}
v_0^{i_1}n+v^{i_2}_0>n, \label{iterative_a}
\end{equation}
\begin{equation}
v_0^{i_1}n(1+n)+v^{i_2}_0n+v^{i_3}_0>n(1+n)+n, \label{iterative_b}
\end{equation}
\begin{equation*}
\vdots
\end{equation*}
\begin{multline}
v_0^{i_1}n\left(1+n\right)^{x-2}+v^{i_2}_0n(1+n)^{x-3}+\ldots+v_0^{i_{x-1}}n+v^{i_x}_0>\\ n\left(1+n\right)^{x-2}+n(1+n)^{x-3}+\ldots+n. \label{iterative_c}
\end{multline}
The right-hand side (RHS) of the above conditions do not depend on the sequence in which the nodes are permanently repaired. Consider the left-hand side (LHS) of the above conditions. In condition \eqref{iterative_a}, the LHS would be the largest when node $i_1$ has the largest initial health (as coefficients corresponding to $v_0^{i_1}$ and $v_0^{i_2}$ are $n$ and 1, respectively). In condition \eqref{iterative_b}, the LHS would be the largest when node $i_1$ has the largest initial health and node $i_2$ has the second largest initial health (as coefficients corresponding to $v_0^{i_1},v_0^{i_2},v_0^{i_3}$ are $n(1+n),n,1$, respectively). Proceeding in this manner until the last condition (equation \eqref{iterative_c}), we see that the LHS would be largest when $i_1$ is the node with largest initial health, $i_2$ is the node with the second largest initial health and so on. Thus, if we define a non-jumping sequence $V$ that targets the nodes of set $\mathcal{Z}$ in decreasing order of their initial health values, it would also permanently repair $x$ nodes and hence will be optimal (since it permanently repairs the same number of nodes as the optimal sequence $U$). Consider another non-jumping sequence $W$ that targets the top $x$ nodes with the largest initial health values from the $N$ nodes. Then, the sequence $W$ would also permanently repair $x$ nodes. This is because each node in sequence $W$ has a higher initial health value (or at least equal) to the corresponding node in sequence $V$ and thus sequence $W$ satisfies the conditions \eqref{iterative_a}-\eqref{iterative_c}. Thus, the policy of targeting the nodes in decreasing order of their initial health values would also permanently repair $x$ nodes, and hence is optimal.  
\end{IEEEproof}
\begin{remark}
Theorem~\ref{thm:nonjumping_optimal_notallfixed} shows that non-jumping policies are optimal when $\Delta_{dec} \ge \Delta_{inc}$. Furthermore, Theorem~\ref{thm:multiplicityconditions_notfixall} shows that under certain conditions on the initial health values, repair/deterioration rates and weights, repairing the nodes in decreasing order of their initial health is optimal. \textbf{Equivalently, under the conditions given in these theorems, the optimal sequence is a feedback policy that targets the healthiest node at each time step.}
\end{remark}
The above theorem relied on the initial health values and rates of repair/deterioration being such that each node requires an integer number of time steps to be permanently repaired (allowing the ceiling functions in the characterization of the number of time steps to be dropped). When the health values and rates do not satisfy those conditions, we provide an example to show that the policy of targeting the nodes in decreasing order of their initial health values need not be optimal. 
\begin{example}
Consider $\Delta_{dec}=0.7, \Delta_{inc}=0.6$, and two nodes having equal weights with initial health values  $v_0^{1}=0.95, v_0^{2}=0.6$. If the node with the largest initial health (i.e., node 1) is first targeted, then node 2 fails by the time the entity reaches it. However, if the node with the lowest initial health (i.e., node 2) is first targeted before targeting node 1 then it is possible to permanently repair both the nodes. Thus, when the conditions of Theorem \ref{thm:multiplicityconditions_notfixall} are not satisfied then targeting the nodes in decreasing order of health values might not be the optimal policy.
\end{example}

We also give an example to show that the policy that targets the healthiest node at each time step may not be optimal when the deterioration and repair rates are not homogeneous. 
\begin{example}
Consider two nodes with equal weights, $v_0^1=0.9, v_0^2=0.4$, $\Delta_{dec}^1=0.6,\Delta_{dec}^2=0.6$, $\Delta_{inc}^1=0.1$, and $\Delta_{inc}^2=0.6$. If the policy of targeting the healthiest node at each time step is followed then node 2 fails by the time the entity reaches it. However, if we follow the non-jumping sequence that first permanently repairs the least healthy node (i.e., node 2), then it is possible to permanently repair both of the nodes. 
\end{example}
We now give an example to show that the policy that targets the healthiest node at each time step may not be optimal when weights are not homogeneous. 
\begin{example} \label{exmp:most_healthy_not_optimal_weighted}
Consider two nodes such that $v_0^1=0.5$, $v_0^2=0.4$, $w^1=1$, $w^2=2$, and homogeneous rates $\Delta_{dec}=\Delta_{inc}=0.1$. If the policy of targeting the healthiest node (i.e., node 1) at each time step is followed then a reward of $1$ is obtained; however, if node 2 is first targeted and permanently repaired, then a reward of $2$ is obtained.  
\end{example}
Characterizing the optimal policy in the above cases is an avenue for future research. However, when the weights are heterogeneous (but the rates are homogeneous) we will next show that the policy that permanently repairs the largest number of nodes also returns an approximately optimal solution to Problem~\ref{problem}.  

\subsection{An approximately optimal policy for heterogeneous weights and homogeneous rates} \label{sec:approx_weights}
We will start with the following general result, relating the optimal sequence for Problem~\ref{problem} to the optimal sequence for permanently repairing the largest number of nodes (i.e., corresponding to the case where all weights are the same). 

\begin{theorem} \label{thm:wmax-approx}
Let there be $N(\geq2)$ nodes with initial health values $v_0 = \{v_0^j\}$ and weights $w = \{w^j\}$. Let $w_{min} = \min_{j}w^j$ and $w_{max}=\max_j w^j$. 
Let $U$ be the optimal sequence for Problem~\ref{problem}, and let $V$ be a control sequence that permanently repairs the largest number of nodes.  
Then, 
$\frac{J(v_0, w, U)}{J(v_0, w, V)}\le\frac{w_{max}}{w_{min}}$, where $J(\cdot, \cdot, \cdot)$ is the reward function defined in Definition~\ref{def:reward}.
\end{theorem}

\begin{IEEEproof}
As defined in Definition~\ref{def:reward}, let the set of nodes permanently repaired by the sequence $V$ be denoted by $\mathcal{M}(v_0, V)\subseteq \mathcal{V}$, and suppose it contains $x$ nodes.  Then, the reward that is obtained by the policy $V$ satisfies $J(v_0, w, V) = \sum_{j\in \mathcal{M}(v_0, V)} w^{j}\ge xw_{min} $.  Let the number of nodes permanently repaired by the optimal sequence $U$ be $y$. Then, the reward that is obtained by the optimal sequence would satisfy $J(v_0, w, U) = \sum_{j \in \mathcal{M}(v_0, U)}w^j\le yw_{max}\le xw_{max}$ as $y\le x$. Therefore, the ratio of the reward obtained by the optimal sequence to the reward obtained by the sequence that permanently repairs the maximum number of nodes satisfies $\frac{J(v_0, w, U)}{J(v_0, w, V)} \le \frac{xw_{max}}{xw_{min}}=\frac{w_{max}}{w_{min}}$.
\end{IEEEproof}

Note that the above result holds regardless of the weights and rates. We now obtain the following result pertaining to instances of Problem~\ref{problem} with homogeneous rates and heterogeneous weights.

\begin{corollary} \label{cor:wmax-approx_healthypolicy}
Let there be $N(\geq2)$ nodes with initial health values $v_0 = \{v_0^j\}$ and weights $w = \{w^j\}$. Let $w_{min} = \min_{j}w^j$ and $w_{max}=\max_j w^j$. 
For all $j\in\{1,\ldots,N\}$, suppose  $\Delta_{inc}^j=\Delta_{inc}$, $\Delta_{dec}^j=\Delta_{dec}$ and $\Delta_{dec}\ge \Delta_{inc}$. Suppose $\Delta_{dec} = n\Delta_{inc}$, where $n$ is a positive integer. Also, for each node $j\in\{1,\ldots,N\}$, suppose there exists a positive integer $m_j$ such that $1-v^{i_j}_0 =m_j \Delta_{inc}$. Then, the policy that targets the healthiest node at each time step provides a reward that is within a factor $\frac{w_{max}}{w_{min}}$ of the optimal reward.
\end{corollary}
The proof of this corollary follows directly from Theorems \ref{thm:multiplicityconditions_notfixall} and \ref{thm:wmax-approx}.

\section{Optimal Control Policies for $\Delta_{dec}^j < \Delta_{inc}^j$}
We now turn our attention to the case where $\Delta_{dec}^j < \Delta_{inc}^j$ for one or more $j \in \{1,\ldots,N\}$.
First, we define the concept of a \textbf{modified health value}. 
\begin{definition}
The modified health value of a node $j$ at time $t$ is the health value minus the rate of deterioration, i.e., $v_t^j-\Delta_{dec}^j$.
\end{definition}

Note that modified health value of a node is allowed to be negative, unlike the health value.  We start with the following general result.

\begin{lemma} \label{lemma3_a}
Let there be $N(\geq 2)$ nodes. Then, for  $z \in \{1, 2, \ldots, N\}$, there exists a sequence that permanently repairs $z$ nodes only if 
there exists a set $\{i_1,\ldots,i_z\}\subseteq \mathcal{V}$ such that 
\begin{equation}
	v_0^{i_j}>(z-j)\Delta_{dec}^{i_j}, \hspace{3mm} \forall j\in \{1,\ldots,z\}. \label{eq:Deltalesslemma_a}
	\end{equation}
\end{lemma}

\begin{IEEEproof}
Suppose there exists a sequence that permanently repairs $z$ nodes. At each time step $t$, use $\mathcal{C}_t$ to denote the set of nodes that have not been targeted at least once by the entity prior to $t$. Note that $\mathcal{C}_0\supseteq \mathcal{C}_1\supseteq \ldots\supseteq \mathcal{C}_{z-1}$. At $t=0$, $|\mathcal{C}_t|=N$ where $|\mathcal{C}_t|$ denotes the cardinality of set $\mathcal{C}_t$. At time $t=1, |\mathcal{C}_{t}|= N-1$ as there are $N-1$ nodes that have not been targeted by the entity at least once. Each node $k$ belonging to the set $\mathcal{C}_1$ should have initial health value larger than $\Delta_{dec}^k$ to survive until $t=1$. At $t=2, |\mathcal{C}_{t}|\ge N-2$ as there are  at least $N-2$ nodes that have not been targeted by the entity at least once. Each node $k$ belonging to the set $\mathcal{C}_2$ should have initial health value larger than $2\Delta_{dec}^k$ to survive until $t=2$. Repeating this argument for the next $z-3$ time steps proves that there must be a permutation $(i_1,\ldots,i_z)$ of nodes that satisfies the conditions \eqref{eq:Deltalesslemma_a} in order for $z$  nodes to eventually be permanently repaired. 
\end{IEEEproof}

Note that \eqref{eq:Deltalesslemma_a} represents necessary conditions that need to be satisfied by {\it any} sequence that permanently repairs all the nodes in the set $\{i_1,\ldots,i_z\}$, regardless of the rates of repair and deterioration. We now provide the following result for the case when the rates of repair are significantly larger than the rates of deterioration.

\begin{lemma} \label{lemma3_b}
Let there be $N(\geq 2)$ nodes. Let $z\le N$ and suppose there exists a set $\{i_1,\ldots,i_z\}\subseteq \mathcal{V}$ such that \eqref{eq:Deltalesslemma_a} holds.
Suppose $\Delta_{inc}^{i_j}>(z-1)\Delta_{dec}^{i_j}, \forall j \in\{1,\ldots,z\}$ and $\Delta_{inc}^{i_j}>\sum_{k\in \{1,\ldots,z\}\setminus j}\Delta_{dec}^{i_k}, \forall j\in\{1,\ldots,z\}$. Then, the sequence that targets the node with the least modified health in the set $\{i_1,\ldots,i_z\}$ at each time step will permanently repair all the nodes of the set $\{i_1,\ldots,i_z\}$.
\end{lemma}

\begin{IEEEproof}
	Suppose there is a set $\{i_1,\ldots,i_z\}$ such that \eqref{eq:Deltalesslemma_a} holds. There are $z$ possible cases depending upon which node in the set $\{i_1,\ldots,i_z\}$ has the lowest initial modified health. The first case is when node $i_z$ has the lowest initial modified health in the set $\{i_1,\ldots,i_z\}$, i.e., $v_0^{i_z}-\Delta_{dec}^{i_z}=\min_{j\in \{1,\ldots,z\}} \{v_0^{i_j}-\Delta_{dec}^{i_j}\}$. After the completion of the first time step, if node $i_z$ does not get permanently repaired, the health values of the nodes are given by 
	\begin{align}
	v^{i_z}_{1} &= v_0^{i_z}+\Delta_{inc}^{i_z}>(z-1)\Delta_{dec}^{i_z}, \\
	v^{i_j}_{1} &= v_0^{i_j}-\Delta_{dec}^{i_j}>(z-1-j)\Delta_{dec}^{i_j}, \label{eq:inc-larger-necc_suff} \\& \forall j\in\{1,\ldots,z-1\}, \nonumber
	\end{align}
where the inequality in \eqref{eq:inc-larger-necc_suff} comes from \eqref{eq:Deltalesslemma_a}. Thus, there exists a permutation $(i'_1,\ldots,i'_z)=(i_z,i_1,i_2,\ldots,i_{z-1})$ that satisfies the conditions \eqref{eq:Deltalesslemma_a} at time $t=1$. However, if node $i_z$ gets permanently repaired after the completion of the first time step, then $v^{i_z}_{1}=1$ and the health values of nodes $\{i_1,\ldots,i_{z-1}\}$ are given by \eqref{eq:inc-larger-necc_suff}. Thus, there exists a permutation $(i'_1,\ldots,i'_{z-1})=(i_1,i_2,\ldots,i_{z-1})$ that satisfies the conditions \eqref{eq:Deltalesslemma_a} (with $z$ replaced by $z-1$) at time $t=1$ along with $v^{i_z}_{1}=1$. We now consider the second case, when $v_0^{i_{z-1}}-\Delta_{dec}^{i_{z-1}}=\min_{j\in\{1,\ldots,z\}} \{v_0^{i_j}-\Delta_{dec}^{i_j}\}$, i.e., node $i_{z-1}$ has the lowest initial modified health in the set $\{i_1,\ldots,i_{z}\}$. Then, after the completion of the first time step, if node $i_{z-1}$ does not get permanently repaired, the health values of the nodes are given by 
	\begin{align}
	v^{i_{z-1}}_{1} &= v_0^{i_{z-1}}+\Delta_{inc}^{i_{z-1}}>(z-1)\Delta_{dec}^{i_{z-1}} , \\
	v^{i_z}_{1} &= v_0^{i_z}-\Delta_{dec}^{i_z}> v_0^{i_{z-1}}-\Delta_{dec}^{i_{z-1}}>0, \label{eq:usingminrelation} \\
	v^{i_j}_{1} &= v_0^{i_j}-\Delta_{dec}^{i_j}>(z-1-j)\Delta_{dec}^{i_j}, \label{eq:healthvalues_1toN-2} \\ &\forall j\in\{1,\ldots,z-2\}. \nonumber
	\end{align}
	Note that the first inequality in condition \eqref{eq:usingminrelation} holds as $v_0^{i_{z-1}}-\Delta_{dec}^{i_{z-1}}=\min_{j\in \{1,\ldots,z\}} \{v_0^{i_j}-\Delta_{dec}^{i_j}\}$. The second inequality in condition \eqref{eq:usingminrelation} holds from \eqref{eq:Deltalesslemma_a}.
	Thus, the nodes of the set $\{i_1,\ldots,i_z\}$ satisfy \eqref{eq:Deltalesslemma_a}, but with the indices reordered. However, if node $i_{z-1}$ gets permanently repaired after the completion of the first time step, then $v^{i_{z-1}}_{1}=1$ and the health values of nodes $\{i_z,i_1,i_2,\ldots,i_{z-2}\}$ are given by \eqref{eq:usingminrelation} and \eqref{eq:healthvalues_1toN-2}. Thus, $z-1$ nodes satisfy \eqref{eq:Deltalesslemma_a} (with $z$ replaced by $z-1$) along with $v^{i_{z-1}}_{1}=1$ after the completion of first time step. The remaining $z-2$ cases similarly follow and are therefore omitted.	
	 Thus, at any time step, if there are $x (\le z)$ nodes that satisfy the conditions in equation \eqref{eq:Deltalesslemma_a} (with $z$ replaced by $x$), and $z-x$ nodes that are permanently repaired, then there will be a permutation of $y(\le x)$ nodes that satisfies the conditions \eqref{eq:Deltalesslemma_a} (with $z$ replaced by $y$) and $z-y$  nodes that are permanently repaired at the start of the next time step. Therefore, no node that belongs to the set $\{i_1,\ldots,i_z\}$ would have health becoming zero at any time. Furthermore, if a node $i_j$, where $j \in \{1,\ldots,z\}$, is targeted by the entity at a time step and it does not get permanently repaired then the average health of the nodes in the set $\{i_1,\ldots,i_z\}$ increases by at least $\frac{\Delta_{inc}^{i_j}-\sum_{k\in\{1,\ldots,z\}\setminus j} \Delta_{dec}^{i_k}}{z}$. Note that $\frac{\Delta_{inc}^{i_j}-\sum_{k\in\{1,\ldots,z\}\setminus j} \Delta_{dec}^{i_k}}{z} >0$ as $\Delta_{inc}^{i_j} >\sum_{k\in\{1,\ldots,z\}\setminus j} \Delta_{dec}^{i_k}, \quad \forall j \in \{1,\ldots,z\}$. So, at each time step, either the increase in average health of the nodes in the set $\{i_1,\ldots,i_z\}$ is positive, or a node gets permanently repaired, or both. Therefore, all the nodes of the set $\{i_1,\ldots,i_z\}$ would eventually be permanently repaired.
\end{IEEEproof}
\begin{remark}
Note that the conditions on the deterioration and repair rates provided in Lemma \ref{lemma3_b} are a function of the particular set of $z$ nodes satisfying \eqref{eq:Deltalesslemma_a}; however, a stronger, but set independent, sufficient condition for the policy given in Lemma \ref{lemma3_b} to repair all the nodes would be $\Delta_{inc}^j>(N-1)\Delta_{dec}^j, \forall j \in \{1,\ldots,N\}$ and $\Delta_{inc}^j >\sum_{k\in\{1,\ldots,N\}\setminus j} \Delta_{dec}^k, \quad \forall j \in \{1,\ldots,N\}$. 
\end{remark}

We will use the above results to show that the optimal policy to solve Problem~\ref{problem} is to target the node with the least modified health in a particular subset of nodes 
at each time step, under certain conditions on the rates of repair and deterioration. This will then show that non-jumping policies are no longer necessarily optimal when $\Delta_{dec}^{j} < \Delta_{inc}^{j}$ for one or more $j\in\{1,\ldots,N\}$.

To derive this optimal policy, we start by presenting Algorithm \ref{alg:create_optimal_set}, which generates a subset $\mathcal{Z}$ from the set of all nodes $\mathcal{V}$. Step 1 of the algorithm outputs a number $x$, which is the largest number such that there exists a set $\{i_1,\ldots,i_x\} \subseteq \mathcal{V}$ satisfying \eqref{eq:Deltalesslemma_a} when $z$ is replaced by $x$ (as we will prove below). Next, in Step 2, a subset $\mathcal{Z}$ of the set  $\mathcal{V}$ is created such that $\mathcal{Z}$ is the set of $x$ nodes with the largest sum of weights while ensuring that the initial health values of the nodes in the set $\mathcal{Z}$ satisfy \eqref{eq:Deltalesslemma_a} when $z$ is replaced by $x$.
 
\begin{algorithm}   \caption{Generation of set $\mathcal{Z}$} \label{alg:create_optimal_set}
Let there be $N (\geq 2)$ nodes. 
  \begin{algorithmic}[1] 
    \State {\bf Computing the largest number $x$ such that there exists a set $\{i_1,\ldots,i_x\} \subseteq \mathcal{V}$ satisfying \eqref{eq:Deltalesslemma_a} when $z$ is replaced by $x$.} 
    First, compute $\left \lceil \frac{v_0^j}{\Delta_{dec}^j} \right \rceil$ for each node $j$. Then, set $x=0$ and let $\mathcal{Y} = \mathcal{V}$ be the set of all $N$ nodes. Then, repeat the following until the termination criterion is satisfied.
    \begin{itemize}
        \item If there is no node $j$ in the set $\mathcal{Y}$ such that $ \left \lceil \frac{v_0^j}{\Delta_{dec}^j} \right \rceil > x$, then terminate this step. Otherwise, let node $j \in \mathcal{Y}$ be the node with the lowest value of $\left \lceil \frac{v_0^j}{\Delta_{dec}^j} \right \rceil$ that satisfies $ \left \lceil \frac{v_0^j}{\Delta_{dec}^j} \right \rceil > x$ among all nodes in $\mathcal{Y}$. Remove node $j$ from the set $\mathcal{Y}$ and set $x=x+1$.   
    \end{itemize}
    \State {\bf Creating a set $\mathcal{Z}$ consisting of $x$ nodes.} Let $\mathcal{W}=\mathcal{V}$ be the set of all $N$ nodes, and let $\mathcal{Z}= \emptyset$. Among all nodes $j$ in $\mathcal{W}$ whose initial health values are larger than $(x-1)\Delta_{dec}^j$, remove the one whose weight is largest and add it to $\mathcal{Z}$.  
    Next, among all nodes $j$ in $\mathcal{W}$ whose initial health values are larger than $(x-2)\Delta_{dec}^j$, remove the one whose weight is largest and add it to $\mathcal{Z}$.  
    Continue in this way until $x$ nodes have been added to $\mathcal{Z}$.
  \end{algorithmic} 
\end{algorithm}

 \begin{remark}
  Note that Algorithm~\ref{alg:create_optimal_set} has polynomial time complexity. Specifically, Step 1 involves computing $\left \lceil \frac{v_0^j}{\Delta_{dec}^j} \right \rceil$ for each node $j$, which takes at most $O(N)$ operations, and then performing min operations over an $O(N)$ array at most $N$ times. In Step 2, every iteration that involves choosing a node for set $\mathcal{Z}$ takes at most $O(N)$ operations (because it involves performing a max operation over an $O(N)$ array) and the maximum size of the set $\mathcal{Z}$ is $N$.
 \end{remark}
 
 We will now show that it is optimal to only target the nodes of set $\mathcal{Z}$ generated by Algorithm~\ref{alg:create_optimal_set} in order to solve Problem~\ref{problem}.  We first prove that Step 1 does indeed find the largest number $x$ such that there exists a set $\{i_1,\ldots,i_x\} \subseteq \mathcal{V}$ satisfying \eqref{eq:Deltalesslemma_a} when $z$ is replaced by $x$.  
 
 \begin{lemma} \label{lem:max_nodes_x_increasesufflarger}
Let there be $N(\ge 2)$ nodes. The value of $x$ that is computed in Step 1 of Algorithm \ref{alg:create_optimal_set} is the largest number such that there exists a set $\{i_1,\ldots,i_x\}\subseteq \mathcal{V}$ satisfying \eqref{eq:Deltalesslemma_a} when $z$ is replaced by $x$.
 \end{lemma}

\begin{IEEEproof}
We prove this result through contradiction. Suppose the value of $x$ that is computed in Step 1 of Algorithm \ref{alg:create_optimal_set} is not the largest number such that there exists a set $\{i_1,\ldots,i_x\}\subseteq \mathcal{V}$ satisfying \eqref{eq:Deltalesslemma_a} when $z$ is replaced by $x$. Then, there exists a set $\{i_1,\ldots,i_y\}\subseteq \mathcal{V}$ of size $y(>x)$ satisfying
 \begin{equation}
    v_0^{i_j}>(y-j)\Delta_{dec}^{i_j}, \quad \forall j \in \{1,\ldots,y\}. \label{eq:nec_cond_increaselarger_contra}
\end{equation}
Assume without loss of generality that these nodes are ordered such that
\begin{equation}
    \left\lceil\frac{v_0^{i_1}}{\Delta_{dec}^{i_1}}\right\rceil \ge \left\lceil\frac{v_0^{i_2}}{\Delta_{dec}^{i_2}}\right\rceil \ge \cdots \ge \left\lceil\frac{v_0^{i_y}}{\Delta_{dec}^{i_y}}\right\rceil.
    \label{eq:opt_number_ordering}
\end{equation}
Note that by \eqref{eq:nec_cond_increaselarger_contra}, and the ordering given in \eqref{eq:opt_number_ordering}, these quantities must satisfy
\begin{equation}
    \left\lceil\frac{v_0^{i_1}}{\Delta_{dec}^{i_1}}\right\rceil > y-1, \enspace  \left\lceil\frac{v_0^{i_2}}{\Delta_{dec}^{i_2}}\right\rceil > y-2, \enspace \ldots, \enspace  \left\lceil\frac{v_0^{i_y}}{\Delta_{dec}^{i_y}}\right\rceil > 0.
    \label{eq:opt_number_condition}
\end{equation}
Now, under the above conditions, we compute the value of $x$ in Step 1 of Algorithm \ref{alg:create_optimal_set}. At the first iteration of Step 1, we have $\mathcal{Y} = \mathcal{V}$ and $x = 0$.  By \eqref{eq:opt_number_condition}, there is at least one node $j \in \mathcal{Y}$ such that $\left\lceil\frac{v_0^{j}}{\Delta_{dec}^{j}}\right\rceil > 0$; for example, $i_y$ satisfies this condition.  Thus, Step 1 does not terminate at this iteration.  Let $k_1 \in \mathcal{Y}$ be the node selected by Step 1, i.e., over all nodes $j \in \mathcal{Y}$ that have $\left\lceil\frac{v_0^{j}}{\Delta_{dec}^{j}}\right\rceil > 0$, $k_1$ has the smallest such ratio.  Note that $k_1 \notin \{i_1, i_2, \ldots, i_{y-1}\}$ by the ordering in \eqref{eq:opt_number_ordering}.

In the second iteration of Step 1, we have $\mathcal{Y} = \mathcal{V}\setminus\{k_1\}$ and $x = 1$.  By \eqref{eq:opt_number_condition}, there is at least one node $j \in \mathcal{Y}$ such that $\left\lceil\frac{v_0^{j}}{\Delta_{dec}^{j}}\right\rceil > 1$; for example, $i_{y-1}$ satisfies this condition. Thus, Step 1 does not terminate at this iteration. Let $k_2 \in \mathcal{Y}$ be the node selected by Step 1, i.e., over all nodes $j \in \mathcal{Y}$ that have $\left\lceil\frac{v_0^{j}}{\Delta_{dec}^{j}}\right\rceil > 1$, $k_2$ has the smallest such ratio.  Note that $k_2 \notin \{i_1, i_2, \ldots, i_{y-2}\}$ by the ordering in \eqref{eq:opt_number_ordering}.

Continuing in this way, in the $r$-th iteration of Step 1 (where $2 \le r \le y-1$), we have $\mathcal{Y} = \mathcal{V}\setminus\{k_1, k_2, \ldots, k_{r-1}\}$ and $x = r-1$.  By \eqref{eq:opt_number_condition}, there is at least one node $j \in \mathcal{Y}$ such that $\left\lceil\frac{v_0^{j}}{\Delta_{dec}^{j}}\right\rceil > r-1$; for example, $i_{y-r+1}$ satisfies this condition. Thus, Step 1 does not terminate at the $r$-th iteration. Let $k_r \in \mathcal{Y}$ be the node selected by Step 1, i.e., over all nodes $j \in \mathcal{Y}$ that have $\left\lceil\frac{v_0^{j}}{\Delta_{dec}^{j}}\right\rceil > r-1$, $k_r$ has the smallest such ratio.  Note that $k_r \notin \{i_1, i_2, \ldots, i_{y-r}\}$ by the ordering in \eqref{eq:opt_number_ordering}.

Finally, in the $y$-th iteration of Step 1, we have $\mathcal{Y} = \mathcal{V}\setminus\{k_1, k_2, \ldots, k_{y-1}\}$ and $x = y-1$.  By \eqref{eq:opt_number_condition}, there is at least one node $j \in \mathcal{Y}$ such that $\left\lceil\frac{v_0^{j}}{\Delta_{dec}^{j}}\right\rceil > y-1$; for example, $i_{1}$ satisfies this condition.  Let $k_y \in \mathcal{Y}$ be the node selected by Step 1, i.e., over all nodes $j \in \mathcal{Y}$ that have $\left\lceil\frac{v_0^{j}}{\Delta_{dec}^{j}}\right\rceil > y-1$, $k_y$ has the smallest such ratio. Thus, the variable $x$ gets set to $y$ at the end of this iteration. However, this leads to a contradiction because we assumed that $y>x$. Therefore, the value of $x$ that is computed in Step 1 of Algorithm \ref{alg:create_optimal_set} is the largest number such that there exists a set $\{i_1,\ldots,i_x\}\subseteq \mathcal{V}$ satisfying \eqref{eq:Deltalesslemma_a} when $z$ is replaced by $x$.
\end{IEEEproof}
 
We now come to the main result of this section.
\begin{theorem} \label{thm:decreasesmaller_jumpingallowed_cannotfixall}
	Let there be $N (\geq 2)$ nodes and let $\mathcal{Z}=\{i_1,\ldots,i_{|\mathcal{Z}|}\}$ be the set that is formed by Algorithm \ref{alg:create_optimal_set}, where $|\mathcal{Z}|=x$. Suppose $\Delta_{inc}^{i_j}>(x-1)\Delta_{dec}^{i_j}, \forall j \in \{1,\ldots,x\}$ and $\Delta_{inc}^{i_j}>\sum_{k\in \{1,\ldots,x\}\setminus j}\Delta_{dec}^{i_k}, \forall j\in \{1,\ldots,x\}$. 
	Then, the optimal policy for Problem~\ref{problem} is to target the node with the least modified health value in the set $\mathcal{Z}$ at each time step.
\end{theorem}

\begin{IEEEproof}
Denote the policy that targets the node with the least modified health value in the set $\mathcal{Z}$ at each time step as $U$. Then, by Lemma \ref{lemma3_b}, all the nodes in set $\mathcal{Z}$ are permanently repaired by $U$ as the initial health values of the nodes in set $\mathcal{Z}$ satisfy \eqref{eq:Deltalesslemma_a} when $z$ is replaced by $x$ (because of the way they are selected in Step 2 of Algorithm \ref{alg:create_optimal_set}). Let $V$ be a sequence other than the sequence $U$. Denote the reward obtained by sequences $U$ and $V$ as $a$ and $b$, respectively. Denote the number of nodes that are permanently repaired by sequences $U$ and $V$ as $x$ and $y$, respectively, and let $\mathcal{S}$ be the set of $y$ nodes that are permanently repaired by sequence $V$. Then, $x\ge y$ by Lemma~\ref{lemma3_a} and Lemma~\ref{lem:max_nodes_x_increasesufflarger}. We argue that $a\ge b$. Let $i_j$ be the $j$th node that is added to the set $\mathcal{Z}$ by Step 2 of Algorithm \ref{alg:create_optimal_set}. Denote the nodes of set $\mathcal{Z}$ by $\{i_1,\ldots,i_x\}$, and the nodes of set $\mathcal{S}$ as $\{i'_1,\ldots,i'_y\}$. In particular, the nodes $i'_1,\ldots,i'_y$ are ordered by performing a similar procedure as in Step 2 of Algorithm \ref{alg:create_optimal_set}. That is, among all nodes $j$ in $\mathcal{S}$ whose initial health values are larger than $(y-1)\Delta_{dec}^j$, we denote the one with the largest weight as node $i'_1$. Next, among all nodes $j$ (other than $i_1'$) in $\mathcal{S}$ whose initial health values are larger than $(y-2)\Delta_{dec}^j$, we denote the one with the largest weight as $i'_2$. We continue this until all the nodes $i'_1,\ldots,i'_y$ are defined.  Note that there must exist at least one node whose initial health value satisfies the specified condition at each iteration, since the nodes in set $\mathcal{S}$ must satisfy the necessary condition \eqref{eq:Deltalesslemma_a} (with $z$ replaced by $y$) in order for all to be permanently repaired.

We prove that there exists a one-to-one mapping between every element of set $\mathcal{S}= \{i'_1,\ldots,i'_y\}$ and an element of set $\mathcal{Z}= \{i_1,\ldots,i_x\}$ such that each mapped node in $\mathcal{Z}$ has a weight that is at least as large as its paired node in $\mathcal{S}$, implying $a\ge b$ (note that it is possible to define such a mapping because $x\ge y$). Let the set of mapped nodes be denoted by $\mathcal{Z^*} = \{i^*_1,\ldots,i^*_y\} \subseteq \mathcal{Z}$. We create the set $\mathcal{Z^*}$ as follows. Node $i^*_1$ is the node with largest weight in the set $\{i_1,\ldots,i_{x-y+1}\}$, $i^*_2$ is the node with largest weight in the set $\{i_1,\ldots,i_{x-y+2}\} \setminus i^*_1$, $i^*_3$ is the node with largest weight in the set $\{i_1,\ldots,i_{x-y+3}\} \setminus \{i^*_1,i^*_2\}$, and so on, until all $y$ nodes of set $\mathcal{Z^*}$ have been defined. Next, for all $j \in \{1,\ldots,y\}$, $i^*_j \in \mathcal{Z}^*$ is mapped to $i'_j \in \mathcal{S}$. We will argue that for all $j \in \{1,\ldots,y\}, w^{i^*_j}\ge w^{i'_j}$.

First, note that $i^*_1$ is the node with largest weight among all nodes $j$ in the set $\mathcal{V}$ whose initial health values are larger than $(y-1)\Delta_{dec}^j$ due to the way the nodes $\{i_1,\ldots,i_{x-y+1}\}$ were chosen in Step 2 of Algorithm~\ref{alg:create_optimal_set}. Since $i'_1$ is the node with largest weight among all nodes $j$ in set $\mathcal{S}$ whose initial health values are larger than $(y-1)\Delta_{dec}^j$ and $\mathcal{S}\subseteq \mathcal{V}$, $w^{i^*_1}\ge w^{i'_1}$ holds true. Next, note that the weight of $i^*_2$ satisfies the following: 1) it is at least as large as the second largest weight among all  nodes $j$ in set $\mathcal{V}$ whose initial health values are larger than $(y-1)\Delta_{dec}^{j}$, and 2) it is at least as large as the largest weight among all nodes $j$ in set $\mathcal{V}$ whose initial health values lie in the interval $((y-2)\Delta_{dec}^{j},(y-1)\Delta_{dec}^{j}]$. Similarly, the weight of $i'_2$ satisfies the following: 1) it is at least as large as the second largest weight among all nodes $j$ in set $\mathcal{S}$ whose initial health values are larger than $(y-1)\Delta_{dec}^{j}$ and 2) it is at least as large as the largest weight among all nodes $j$ in set $\mathcal{S}$ whose initial health values lie in the interval $((y-2)\Delta_{dec}^{j},(y-1)\Delta_{dec}^{j}]$. Therefore, $w^{i^*_2}\ge w^{i'_2}$ holds true because $\mathcal{S}\subseteq \mathcal{V}$.  Continuing in this way we can show that for all $j \in \{1,\ldots,y\}, w^{i^*_j}\ge w^{i'_j}$.

Thus, since the total weight of the nodes permanently repaired by sequence $U$ is at least as large as the total weight of the nodes permanently repaired by any other sequence, we see that the sequence that targets the node with the least modified health in the set $\mathcal{Z}$ at each time step is optimal for Problem~\ref{problem}.
\end{IEEEproof}

We now provide an example to illustrate the generation of set $\mathcal{Z}$ and the policy of targeting the node with least modified health value in the set $\mathcal{Z}$ at each time step.

\begin{example} \label{exmp:setZ_optimalpolicy}
Consider three nodes such that $v_0^1=0.3, v_0^2=0.5, v_0^3=0.2$, $w^1=3, w^2=1, w^3=2$, $\Delta_{dec}^1=0.4,\Delta_{dec}^2=0.3, \Delta_{dec}^3=0.4$, $\Delta_{inc}^1=0.9$, $\Delta_{inc}^2=0.85$ and $\Delta_{inc}^2=0.95$. The values of $\left \lceil \frac{v_0^j}{\Delta_{dec}^j} \right \rceil $ for nodes 1, 2, and 3 are 1, 2, and 1, respectively. Therefore, the largest number $x$ such that there exists a set $\{i_1,\ldots,i_x\}\subseteq \mathcal{V}$ satisfying \eqref{eq:Deltalesslemma_a} when $z$ is replaced by $x$ from Step 1 of Algorithm \ref{alg:create_optimal_set} is two. In Step 2 of the algorithm, node 2 is first selected for the set $\mathcal{Z}$ because it is the only node whose initial health value is larger than the corresponding deterioration rate. After this, node 1 is added to the set $\mathcal{Z}$ because it has the largest weight among the nodes 1 and 3, both of whose initial health values are positive. Therefore, set $\mathcal{Z}$ contains nodes 1 and 2. By Theorem \ref{thm:decreasesmaller_jumpingallowed_cannotfixall}, the optimal policy is to target the node with the least modified health value in the set $\mathcal{Z}$ at each time step. At time step 0, node 1 has the least modified health value in the set $\mathcal{Z}$ and thus it is targeted in the first time step. Table \ref{table_example_setZ} shows the progression of health values of nodes when the optimal policy is followed.  The optimum reward in this example is thus given by $w^1 + w^2 = 4$.
\end{example}

\begin{table}[h]
	\caption{Health progression when the optimal policy is followed in Example \ref{exmp:setZ_optimalpolicy}.}
	\label{table_example_setZ}
	\begin{center}
		\begin{tabular}{|cccc|}
			\hline
			Time step $(t)$&$v_t^{1}$& $v_t^{2}$ & $v_t^{3}$ \\ \hline
			0&0.3& 0.5 &0.2\\
			1&1& 0.2&0 \\
			2&1& 1 &0\\ \hline
		\end{tabular}
	\end{center}
\end{table}
We now consider a special case of Problem \ref{problem} when the weights are homogeneous across all the nodes, i.e., for all $ j\in \{1,\ldots,N\}$, $w^j=\overline{w}$. We show that in this case, it is not required to generate the set $\mathcal{Z}$ through Algorithm \ref{alg:create_optimal_set} to optimally target the nodes.

\begin{proposition} \label{prop:decreasesmaller_jumpingallowed}
Let there be $N (\geq 2)$ nodes such that for all $j\in \{1,\ldots,N\}$, $w^j=\overline{w}$. Suppose $\Delta_{inc}^j>(N-1)\Delta_{dec}^j, \forall j \in \{1,\ldots,N\}$ and $\Delta_{inc}^j >\sum_{k\in \{1,\ldots,N\}\setminus j} \Delta_{dec}^k, \quad \forall j \in \{1,\ldots,N\}$. Then, the policy that targets the node with the least modified health (and that has not permanently failed) at each time step is optimal.
\end{proposition}

\begin{IEEEproof}
Consider an optimal sequence $U$, and let $x(\le N)$ be the number of nodes permanently repaired by that sequence. Denote the set of $x$ nodes as $\mathcal{S}$. By Lemma \ref{lemma3_a}, there exists a permutation $(i_1,\ldots,i_x)$ of the nodes in the set $\mathcal{S}$ such that \eqref{eq:Deltalesslemma_a} is satisfied when $z$ is replaced by $x$. 
Based on the conditions on the repair and deterioration rates assumed in the proposition, the sequence $V$ that targets the node with the least modified health at each time step in $\mathcal{S}$ permanently repairs all of the nodes in $\mathcal{S}$ by Lemma \ref{lemma3_b}.

Let $\mathcal{B}_0$ be the set of nodes that satisfies \eqref{eq:Deltalesslemma_a} (with $z$ replaced by $x$) at time step $0$ and denote the set of nodes that are in the permanent repair state at time step $0$ as $\mathcal{B}'_0$.
Then, $\mathcal{B}_0$ is the set $\mathcal{S}$ and $\mathcal{B}'_0=\emptyset$.
Consider the policy in which the entity targets the node in $\mathcal{V}$ with the least modified health value (and that has not permanently failed) at each time step. Then, in the first time step, either the node with the least modified health value from the set $\mathcal{B}_0$ is targeted or a node outside the set $\mathcal{B}_0$ is targeted. If a node from the set $\mathcal{B}_0$ is targeted and at the end of first time step no node gets permanently repaired, then all the nodes from the set $\mathcal{B}_0$ satisfy the conditions \eqref{eq:Deltalesslemma_a} (with $z$ replaced by $x$) and in that case we define the set $\mathcal{B}_1$ to be the same as set $\mathcal{B}_0$, and define $\mathcal{B}'_1=\emptyset$. If a node from the set $\mathcal{B}_0$ is targeted and gets permanently repaired during that time step, then the remaining $x-1$ nodes from the set $\mathcal{B}_0$ satisfy the conditions \eqref{eq:Deltalesslemma_a} when $z$ is replaced by $x-1$ (as argued in the proof of Lemma \ref{lemma3_b}). In that case we define the set $\mathcal{B}_1$ to be the subset of $x-1$ nodes from $\mathcal{B}_0$ that are not permanently repaired, and define $\mathcal{B}'_1$  to be the node that lies in the set $\mathcal{B}_0\setminus \mathcal{B}_1$. Consider the other case in which a node $c$ not belonging to the set $\mathcal{B}_0$ is targeted in the first time step. Then, if node $c$ does not get permanently repaired, the health value of node $c$ after the first time step would be greater than $(x-1)\Delta_{dec}^c$ as $\Delta_{inc}^c>(N-1)\Delta_{dec}^c\ge (x-1)\Delta_{dec}^c$. Also, a set of $x-1$ nodes in the set $\mathcal{B}_0$ would satisfy the following due to conditions \eqref{eq:Deltalesslemma_a} (with $z$ replaced by $x$):
\begin{equation}
v_1^{i_j}=v_0^{i_j}-\Delta_{dec}^{i_j}> (x-j-1)\Delta_{dec}^{i_j}, \hspace{3mm} \forall j\in \{1,\ldots,x-1\}. \label{eq:healthvalues_1tox-1}
\end{equation}
Thus, if node $c$ does not get permanently repaired after the completion of the first time step, then define $\mathcal{B}_1$ to be the set of nodes (consisting of node $c$ and $x-1$ nodes from $\mathcal{B}_0$) that satisfies the conditions \eqref{eq:Deltalesslemma_a} (with $z$ replaced by $x$), and define $\mathcal{B}'_1=\emptyset$. If node $c$ gets permanently repaired after the completion of the first time step, then $v_1^c=1$ and the health values of a set of $x-1$ nodes in the set $\mathcal{B}_0$ satisfy \eqref{eq:healthvalues_1tox-1}. Thus, if node $c$ gets permanently repaired after the end of the first time step, then define $\mathcal{B}_1$  to be the set that consists of $x-1$ nodes that satisfy the conditions \eqref{eq:Deltalesslemma_a} (with $z$ replaced by $x-1$), and define $\mathcal{B}'_1=c$. We can repeat this argument for all the subsequent time steps, noting that at the end of time step $t$, depending on the sequence of nodes that are targeted by the entity, the initial health values of nodes, and deterioration and repair rates, there would always be a set $\mathcal{B}_t$ (of size $x$ or less) that would satisfy the conditions \eqref{eq:Deltalesslemma_a} (with $z$ replaced by $|\mathcal{B}_t|$) and there would be a set $\mathcal{B}'_t$ of size $x-|\mathcal{B}_t|$ consisting of permanently repaired nodes.

Denote the set of all nodes that have health values in the interval $(0,1]$ at the beginning of time step $t$ by $\mathcal{C}_t$ (i.e., $\mathcal{C}_t$ consists of all the nodes except the nodes that are in the permanent failure state at the beginning of time step $t$). Then, $\mathcal{C}_0=\mathcal{V}$ and for all time steps $t (\ge 0)$, $\mathcal{C}_{t+1} \subseteq \mathcal{C}_{t}$. 
Let a node $i_j\in \mathcal{C}_{t}\setminus \mathcal{B}'_t$, where $j\in \{1,\ldots,|\mathcal{C}_{t}\setminus \mathcal{B}'_t|\}$, be targeted by the entity at time step $t$ and assume that it does not get permanently repaired during time step $t$. Recall that $\mathcal{C}_{t+1}$ consists of all the nodes except the nodes that are in permanent failure state at the beginning of time step $t+1$. We now compute the difference in the average health values of the nodes in $\mathcal{C}_{t+1}$ and $\mathcal{C}_t$ as follows:
\begin{align*}
\frac{{\sum_{i_k \in \mathcal{C}_{t+1}}v_{t+1}^{i_k}}}{|\mathcal{C}_{t+1}|} &-\frac{{\sum_{i_k \in \mathcal{C}_{t}}v_{t}^{i_k}}}{|\mathcal{C}_{t}|} \\ 
&\ge \frac{{\sum_{i_k \in \mathcal{C}_{t+1}}v_{t+1}^{i_k}}}{|\mathcal{C}_{t}|}-\frac{{\sum_{i_k \in \mathcal{C}_{t}}v_{t}^{i_k}}}{|\mathcal{C}_{t}|}\\
&=\frac{\sum_{i_k \in \mathcal{C}_{t}}v_{t+1}^{i_k}}{|\mathcal{C}_{t}|}-\frac{{\sum_{i_k \in \mathcal{C}_{t}}v_{t}^{i_k}}}{|\mathcal{C}_{t}|}\\
&=\frac{\sum_{i_k\in \mathcal{C}_t\setminus \mathcal{B}'_t}\left(v_{t+1}^{i_{k}}-v_t^{i_k}\right)}{|\mathcal{C}_t|}\\
&\ge \frac{\Delta_{inc}^{i_j}-\sum_{k\in\{1,\ldots,|\mathcal{C}_{t}\setminus\mathcal{B}'_t| \}\setminus j }\Delta_{dec}^{i_k}}{|\mathcal{C}_t |}.
\end{align*}
The first inequality above is because  $\mathcal{C}_{t+1}\subseteq \mathcal{C}_{t}$ for all time steps $t$, the first equality is because the health value of a node belonging to the set $\mathcal{C}_t\setminus \mathcal{C}_{t+1}$ at the beginning of time step $t+1$ is equal to zero (i.e., the permanent failure state), and the next equality is because all the nodes in the set $\mathcal{B}'_t$ are in permanent repair state for all time steps greater than or equal to $t$. The last inequality is due to the fact that some of the nodes in the set $\mathcal{C}_t\setminus \mathcal{B}'_t$ that fail during time step $t$ may have had  health values less than their corresponding deterioration rates, and thus the decrease in their health during that time step will also be less than their deterioration rate. 
Note that $\frac{\Delta_{inc}^{i_j}-\sum_{k\in\{1,\ldots,|\mathcal{C}_{t}\setminus\mathcal{B}'_t| \}\setminus j }\Delta_{dec}^{i_k}}{|\mathcal{C}_t |}$ is lower bounded by a  positive constant value equal to $ \frac{\Delta_{inc}^{i_j}-\sum_{k\in\{1,\ldots,N\}\setminus j }\Delta_{dec}^{i_k}}{N}$ because of the conditions on the repair and deterioration rates assumed in the proposition. Therefore, for each time step $t(\ge 0)$, either the average health of the nodes in the set $\mathcal{C}_{t+1}$ is larger than the the average health of the nodes in the set $\mathcal{C}_{t}$, or a node from the set $\mathcal{C}_{t}\setminus \mathcal{B}'_{t}$ gets permanently repaired during time step $t$, or both. Thus, $x$ nodes would eventually get permanently repaired because $ |\mathcal{C}_t|\ge x$ (as $\mathcal{B}_t \cup \mathcal{B}'_t\subseteq \mathcal{C}_t$), for all time steps $t$. 
Therefore, if there is an optimal sequence $U$ that permanently repairs $x(\le N)$ nodes then the sequence that targets the node with the least modified health (and that has not permanently failed) at each time step also permanently repairs $x$ nodes. The result thus follows.  
\end{IEEEproof}
It can be seen that optimal control sequences depend on the relationship between $\Delta_{dec}^j$ and $\Delta_{inc}^j$. When the rates and weights are homogeneous across all the nodes and $\Delta_{dec}\geq\Delta_{inc}$, targeting the healthiest node at each time step is the optimal feedback policy (under certain conditions on the initial health values) by Theorems \ref{thm:nonjumping_optimal_notallfixed} and \ref{thm:multiplicityconditions_notfixall}, whereas
targeting the least healthy node at each time step is the optimal feedback policy when $\Delta_{inc}>(N-1)\Delta_{dec}$ by Proposition \ref{prop:decreasesmaller_jumpingallowed} (when the deterioration rates are homogeneous across all the nodes, the node with the least modified health value is equivalent to the node with the least health value). 

While we have identified the optimal policies for the above ranges of repair and deterioriation rates, the characterization of the optimal policy when  $\Delta_{dec}<\Delta_{inc}<(N-1)\Delta_{dec}$ remains open.  In particular, we provide an example to show that the above optimal sequences will generally not be optimal in this range.

\begin{example} \label{example1}
Consider three nodes with homogeneous weights and rates such that $\Delta_{inc} = 0.025$ and $\Delta_{dec} = 0.02$, and therefore $\Delta_{dec}<\Delta_{inc}<(N-1)\Delta_{dec}$. Suppose $ v_0^{1} = 0.8$,  $v_0^{2} = 0.52$ and $v_0^{3} =0.73$. Consider a non-jumping sequence that targets the nodes in the order $(1, 2, 3)$; one can verify that this sequence permanently repairs all the nodes. However, the non-jumping sequence $(1, 3, 2)$ that targets nodes in decreasing order of their health values does not permanently repair all the nodes. Table \ref{table_example1} presents the progression of health values of nodes for the aforementioned sequences. Additionally, consider the sequence that targets the least healthy node at each time step (i.e., the optimal policy under homogeneous rates and weights when $\Delta_{inc}>(N-1)\Delta_{dec}$). This sequence also does not permanently repair all nodes. Table \ref{table_example3} presents the progression of health values of nodes when the least healthy node is targeted at each time step.
\end{example}

\begin{table}[h]
	\caption{Health  progression with non-jumping sequences $\left(1,2,3\right)$ (left) and $\left(1,3,2\right)$ (right) in Example \ref{example1}.}
	\label{table_example1}
	\begin{center}
		\begin{tabular}{|>{\centering}m{1.3cm}>{\centering}m{0.3cm}>{\centering}m{0.3cm}>{\centering}m{0.5cm}|>{\centering}m{1.3cm}>{\centering}m{0.3cm}>{\centering}m{0.3cm}>{\centering}m{0.5cm}|}
			\hline
			Time step $(t)$&$v_t^{1}$& $v_t^{2}$ &$v_t^{3}$&Time step $(t)$&$v_t^{1}$&$v_t^{2}$ & $v_t^{3}$\tabularnewline\hline		
			0&0.8& 0.52 &0.73&0&0.8&0.52 & 0.73\tabularnewline
			8&1& 0.36 &0.57&8&1&0.36 & 0.57\tabularnewline
			34&1& 1 &0.05&26&1&0 & 1\tabularnewline
			72&1& 1 &1& && & \tabularnewline\hline
		\end{tabular}
	\end{center}
\end{table}

\begin{table}[h]
	\caption{Health progression when the least healthy node is targeted at each time step in Example \ref{example1}.}
	\label{table_example3}
	\begin{center}
		\begin{tabular}{|cccc|}
			\hline
			Time step $(t)$&$v_t^{1}$& $v_t^{2}$ &$v_t^{3}$ \\ \hline		
			0&0.8& 0.52 &0.73\\
			1&0.78& 0.545 &0.71\\
			2&0.76& 0.57 &0.69\\
 			3&0.74& 0.595 &0.67\\
			\vdots&\vdots&\vdots&\vdots \\ 
			134&0.01& 0&0.03\\ 
			\vdots&\vdots&\vdots&\vdots \\ \hline
		\end{tabular}
	\end{center}
\end{table}

\section{Simulation results}
In this section, we seek to understand how much better the optimal policy can perform compared to \textit{randomly generated} sequences. In a \textit{randomly generated} sequence, a node is chosen uniformly random from all the nodes that have health values in the interval (0,1) (i.e., the nodes that are not permanently failed or repaired) at each time-step. In these tests, we keep the weights as well as the deterioration and repair rates to be homogeneous. We split our results into two parts: 1) $\Delta_{dec}\ge \Delta_{inc}$, and 2) $\Delta_{dec}<\Delta_{inc}$. 

In the first case, consider $\Delta_{dec}=0.01$, $\Delta_{inc}=0.01$ and 15 nodes that have identical initial health values equal to $0.99$. These parameters satisfy the conditions of Theorem \ref{thm:multiplicityconditions_notfixall}. Therefore, the sequence that targets the nodes in decreasing order of health values is optimal. By simulation, we find that the number of nodes that are permanently repaired by the optimal sequence is equal to 7. For this example, the maximum number of nodes that can be permanently repaired can also be calculated by Lemma \ref{lem:maxnodes} with $n=\frac{\Delta_{dec}}{\Delta_{inc}}=1$. Note that $  \left \lfloor \log_{(1+n)} \left( \frac{n}{\min_j \{\Delta_{dec}^j \}} +1\right)+1 \right \rfloor= \left \lfloor \log_2 \left( \frac{1}{0.01} +1\right)+1 \right \rfloor =7<15$. Thus, the maximum number of nodes that can be permanently repaired is $L = 7$.
 
\begin{figure}[ht]
	\centering
	\includegraphics[width= 0.45\textwidth]{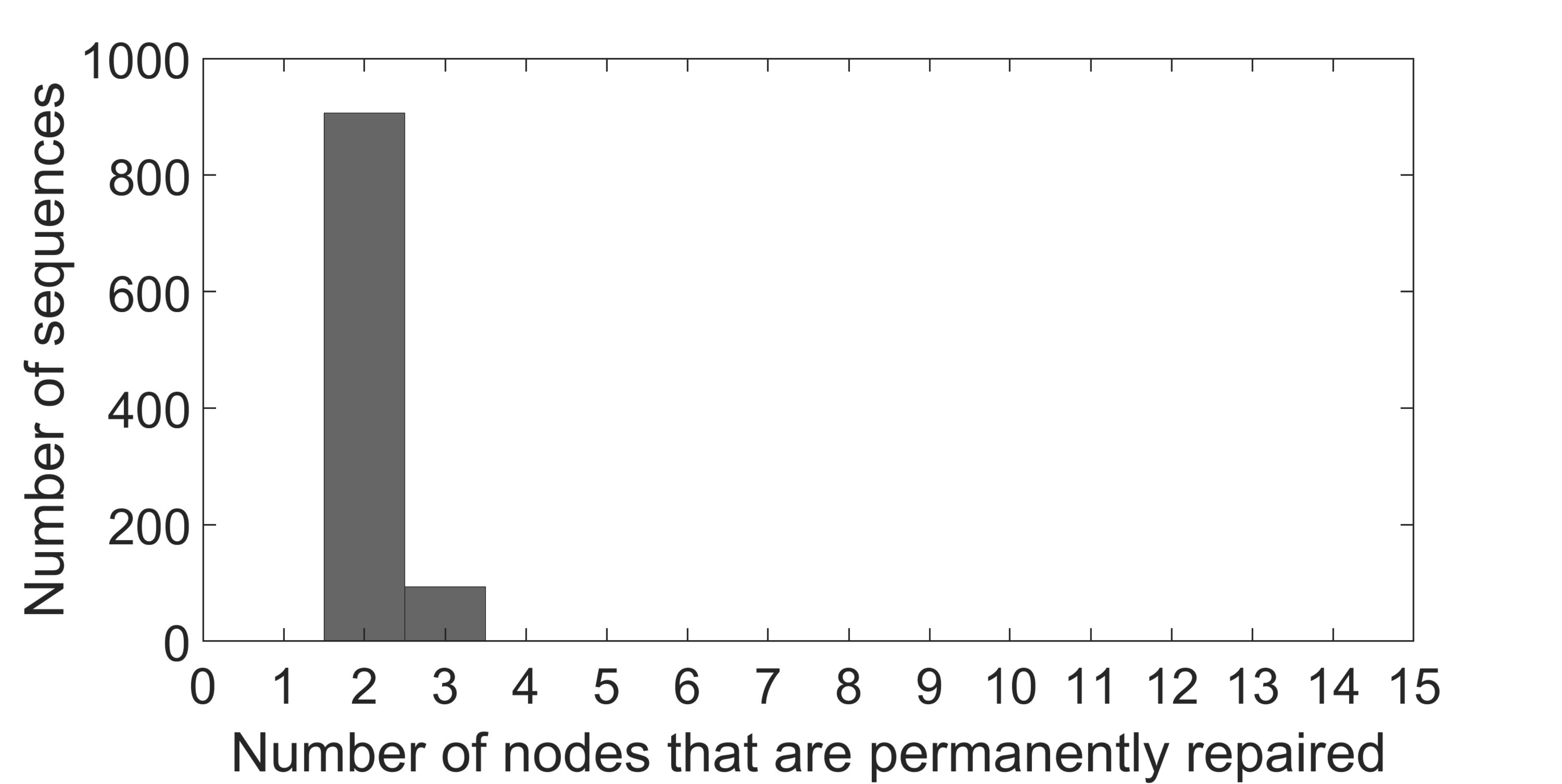}
	\caption{Histogram of number of nodes that are permanently repaired by randomly generated sequences in case 1. }
	\label{fig:sim_dec_larger}
\end{figure}

\begin{figure}[ht]
	\centering
	\includegraphics[width= 0.45\textwidth]{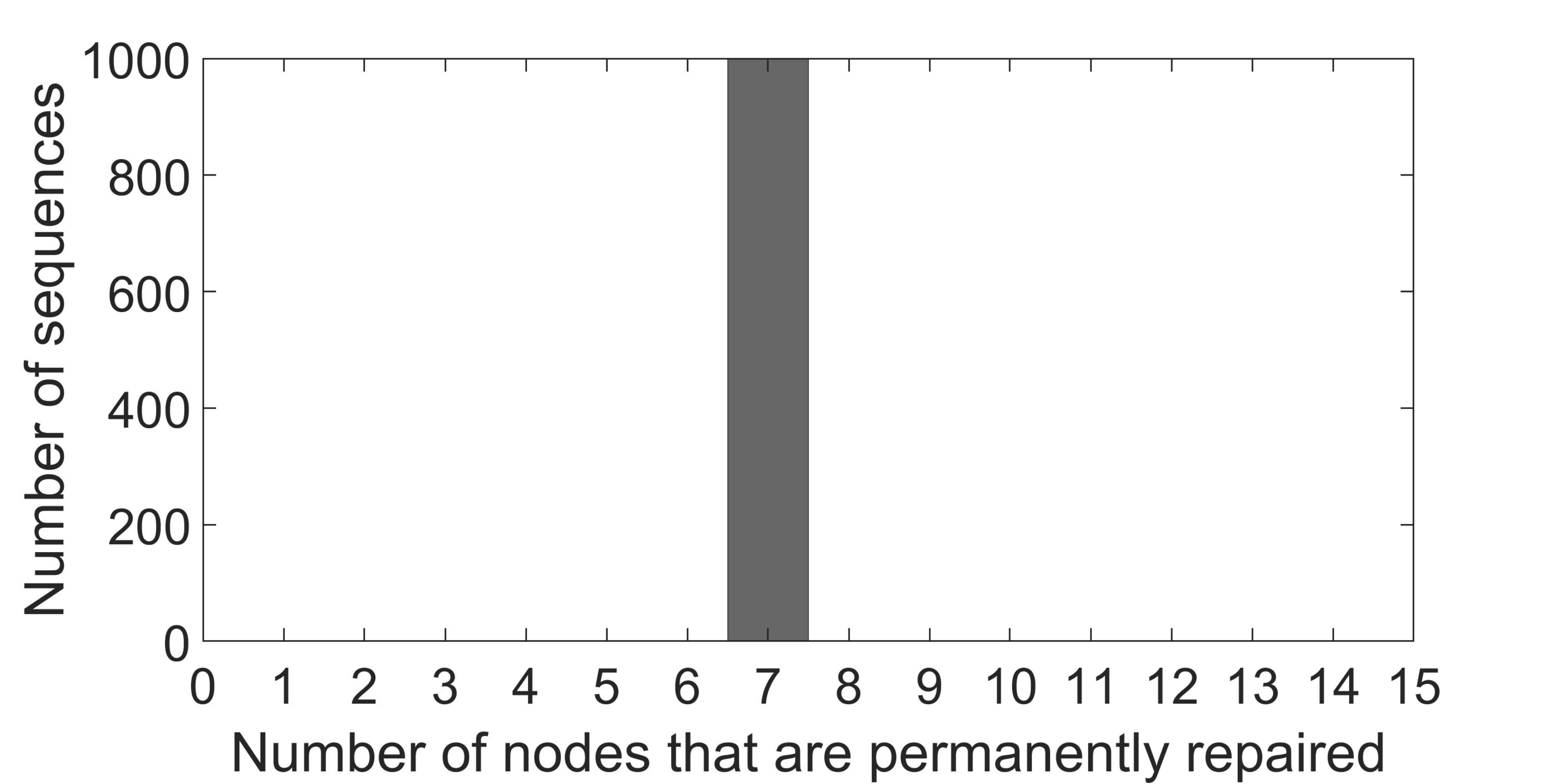}
	\caption{Histogram of number of nodes that are permanently repaired by randomly generated non-jumping sequences in case 1. }
	\label{fig:sim_dec_larger_nojump}
\end{figure}

To compare how much better the optimal policy does than randomly generated sequences, we randomly generated 1000 sequences (without any restriction on jumps) and computed the number of nodes that are permanently repaired by each one. Figure \ref{fig:sim_dec_larger} presents the distribution of nodes that are permanently repaired by the randomly generated sequences. It can be seen that most of the sequences permanently repair two nodes. We also randomly generated non-jumping sequences and plotted the distribution of nodes that are permanently repaired by such sequences in Figure \ref{fig:sim_dec_larger_nojump} (in a randomly generated non-jumping sequence, a node is chosen uniformly random from all the nodes that have health values in the interval (0,1) at the given time and then that node is permanently repaired before  another node is targeted by the entity). It can be seen that all the non-jumping sequences permanently repair 7 nodes; this is due to the initial health values of all the nodes being equal (causing all nodes to be identical in this example). Another important point from Figures \ref{fig:sim_dec_larger} and \ref{fig:sim_dec_larger_nojump} is that non-jumping sequences permanently repair more nodes than general sequences because non-jumping policies are optimal when $\Delta_{dec}\ge \Delta_{inc}$. 

Note that in the aforementioned example, all non-jumping sequences perform equally well. However, this will not always hold true. 
For example, consider a setting with $N (\ge 3)$ nodes. For all $j\in\{1,\ldots,N\}$, let $\Delta_{dec}^j=\Delta_{dec} = \frac{1}{N}$, and  $\Delta_{inc}^j=\Delta_{inc} = \frac{1}{N}$. Out of the $N$ nodes, let there be a set $\mathcal{B}$ with $\left \lfloor \log_{2} \left( N +1\right)+1 \right \rfloor$ nodes that have initial health values equal to $1-\Delta_{inc}=1-\frac{1}{N}$ and a set $\mathcal{C}$ having the remaining nodes with initial health values equal to $\Delta_{inc}=\frac{1}{N}$. Then, the optimal sequence (that targets the nodes in decreasing order of initial health values) permanently repairs at most $\left \lfloor \log_{2} \left( N +1\right)+1 \right \rfloor $ nodes by Lemma \ref{lem:maxnodes}. Therefore, the optimal sequence permanently repairs a subset of nodes in the set $\mathcal{B}$. Note that the first time step at which a node $j\in\mathcal{B}$ reaches the permanent failure state in the optimal sequence is equal to $\frac{v_0^j}{\Delta_{dec}^j}=\frac{1-\frac{1}{N}}{\frac{1}{N}}=N-1$. Also, the number of time steps taken to permanently repair the $i_j$th node in the optimal sequence is equal to $\left(1+\frac{\Delta_{dec}}{\Delta_{inc}}\right)^{j-1}= 2^{j-1}$. Thus, it takes $2^0+2^1+\ldots+2^{x-2}=2^{x-1}-1$ time steps to permanently repair $x$ nodes in the optimal sequence. Therefore, the number of nodes that can be permanently repaired in the optimal sequence in $N-2$ time steps is $x= \left \lfloor \log_{2} \left( N -1\right)+1 \right \rfloor$. Therefore, as $N\rightarrow \infty$, the number of nodes permanently repaired by the optimal sequence goes to infinity.
Consider a non-jumping sequence that first targets one of the nodes of set $\mathcal{C}$. Then, this sequence would only be able to permanently repair one node (as all the other nodes would fail by the time the entity starts targeting them). The probability that a randomly generated non-jumping sequence would start targeting one of the nodes of set $\mathcal{B}$ is $\frac{\left \lfloor \log_{2} \left( N +1\right)+1 \right \rfloor}{N}$. Thus, as $N\rightarrow \infty$, the probability that a randomly generated non-jumping sequence permanently repairs more than one node goes to zero. Therefore, the optimal sequence does infinitely better than a randomly generated non-jumping sequence with probability one as $N\rightarrow \infty$.

\begin{figure}[ht]
	\centering
	\includegraphics[width= 0.45\textwidth]{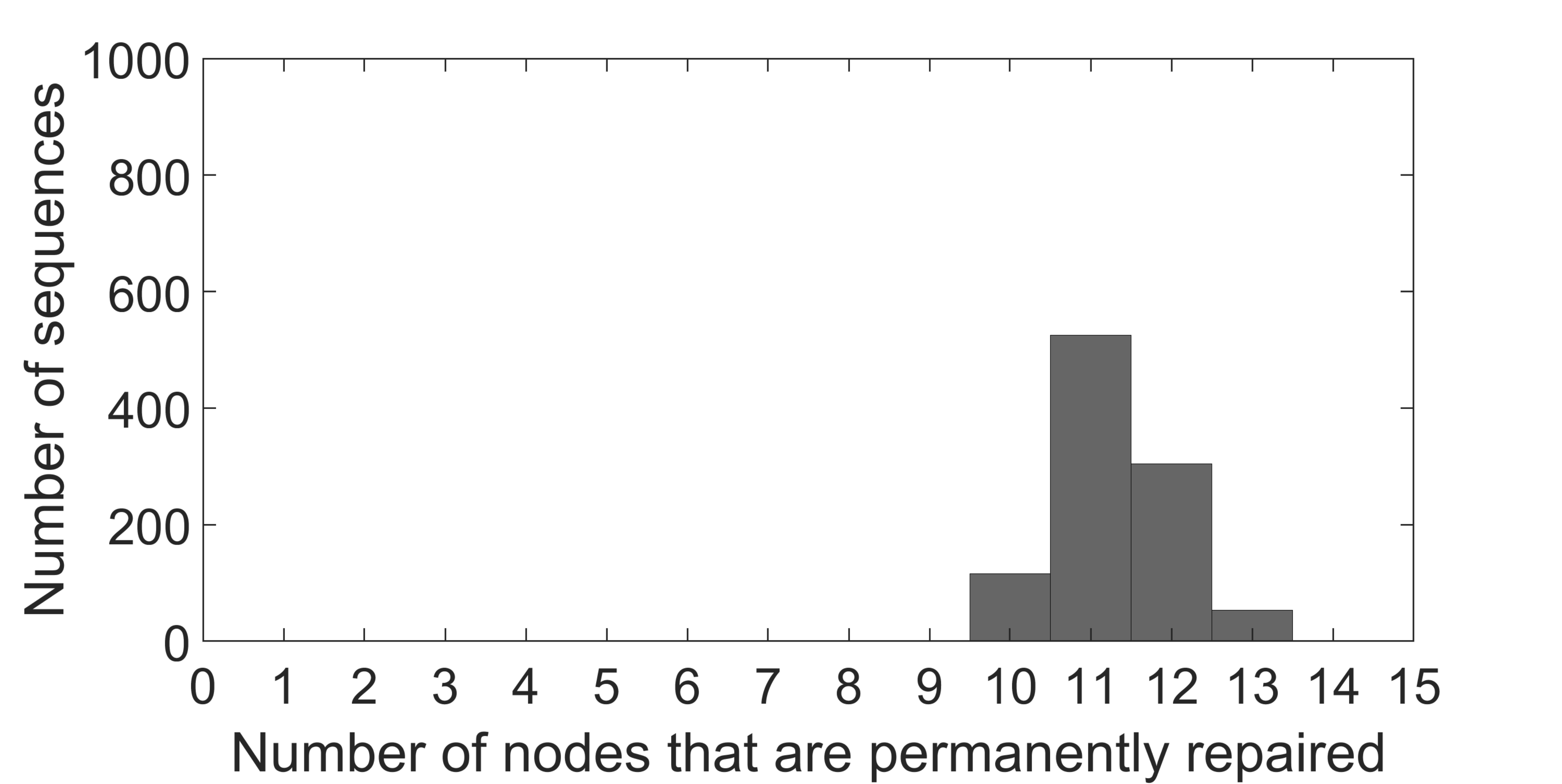}
	\caption{Histogram of number of nodes that are permanently repaired by randomly generated sequences in case 2.}
	\label{fig:sim_inc_larger}
\end{figure}

Next, consider another case where the initial health values of the 15 nodes are equal to $0.05,0.1,\ldots,0.75$. Let $\Delta_{dec}=0.03$  and $\Delta_{inc}=0.75$, so that the condition $\Delta_{inc}>(N-1)\Delta_{dec}$ is satisfied. Thus, the sequence that targets the least healthy node at each time step is optimal (under homogeneous rates and weights) by Proposition \ref{prop:decreasesmaller_jumpingallowed}. By simulating this sequence, we find that the sequence permanently repairs all the 15 nodes. Figure \ref{fig:sim_inc_larger} presents the distribution of nodes that are permanently repaired by randomly generated sequences. It can be seen that the random sequences permanently repair approximately $11$ nodes.  

The aforementioned simulation results show that the number of nodes that are permanently repaired by randomly generated sequences can be significantly different from the optimal sequences, and thereby illustrate the benefit of characterizing and using the optimal policies.

\section{Conclusion}
In this paper, we studied a control problem in a scenario where multiple components' states (or health values) have been reduced following a disruption, and an entity (or controller) aims to maximize the reward of the components whose states are brought to a permanent repair state. This problem comes under the general class of optimal control and scheduling of discrete-time linear switched systems with a non-linear objective function. We 
characterized optimal control policies for specific instances of the problem. We found that the characteristics of the optimal policies depend on the relationship between the rates of repair and deterioration. We showed that when the deterioration and repair rates, as well as the weights associated with the components, are homogeneous, and the deterioration rate is larger than the repair rate, it is optimal to target the healthiest component at each time step (under certain conditions on the initial health values). If the repair rate is sufficiently greater than the deterioration rate for each component, then it is optimal to target the component with the least modified health in a particular subset of components at each time step. 

There are several interesting avenues for future research. Characterizing optimal policies with a constraint on the maximum number of time steps that are available to repair nodes would be important for real-world scenarios (e.g., due to a limited repair budget and other external factors). We believe that the optimal control policies that we characterized when there is no time-constraint can be extended for the case when there is a time-constraint, given that the deterioration rates are larger than the repair rates. However, the characterization of optimal policies when repair rates are larger than deterioration rates remains open for future work. Characterizing optimal policies with non-constant deterioration and repair rates, or with stochastic deterioration and repair rates, is another potential avenue. Also, developing state estimation methods for exact measurement of the health values and the rates of the components will be of interest. Furthermore, incorporating precedence relations between different components into the control decisions also has importance for real-world scenarios.  Finally, one can consider scenarios where the entity can target multiple components simultaneously, or where multiple entities are involved.

\bibliographystyle{IEEEtran}
\bibliography{refs}
\begin{IEEEbiography}[{\includegraphics[width=1in,height=1.25in,clip,keepaspectratio]{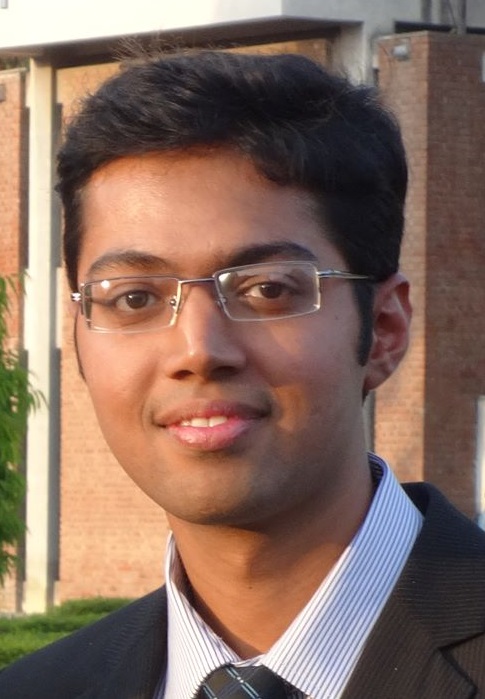}}]{Hemant~Gehlot}
is a graduate student in the Lyles School of Civil Engineering at Purdue University and is pursuing his PhD under the supervision of Dr. Satish V. Ukkusuri and Dr. Shreyas Sundaram. He received his BTech-MTech (dual) degrees from the Indian Institute of Technology Kanpur in 2015. He was a finalist for the Best Student Paper Award at the IFAC Workshop on Distributed Estimation and Control in Networked Systems (NecSys) 2019. He received Essam and Wendy Radwan Graduate Fellowship by Purdue University. His research interests include optimal control and combinatorial optimization.
\end{IEEEbiography}

\begin{IEEEbiography}[{\includegraphics[width=1in,height=1.25in,clip,keepaspectratio]{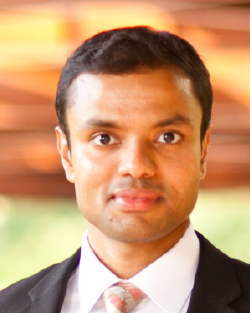}}]{Shreyas~Sundaram}
is an Associate Professor in the School of Electrical and Computer Engineering at Purdue University. He received his MS and PhD degrees in Electrical Engineering from the University of Illinois at Urbana-Champaign in 2005 and 2009, respectively. He was a Postdoctoral Researcher at the University of Pennsylvania from 2009 to 2010, and an Assistant Professor in the Department of Electrical and Computer Engineering at the University of Waterloo from 2010 to 2014. He is a recipient of the NSF CAREER award, and an Air Force Research Lab Summer Faculty Fellowship. At Purdue, he received the Hesselberth Award for Teaching Excellence and the Ruth and Joel Spira Outstanding Teacher Award. At Waterloo, he received the Department of Electrical and Computer Engineering Research Award and the Faculty of Engineering Distinguished Performance Award. He received the M. E. Van Valkenburg Graduate Research Award and the Robert T. Chien Memorial Award from the University of Illinois, and he was a finalist for the Best Student Paper Award at the 2007 and 2008 American Control Conferences. His research interests include network science, analysis of large-scale dynamical systems, fault-tolerant and secure control, linear system and estimation theory, game theory, and the application of algebraic graph theory to system analysis.
\end{IEEEbiography}

\begin{IEEEbiography}[{\includegraphics[width=1in,height=1.25in,clip,keepaspectratio]{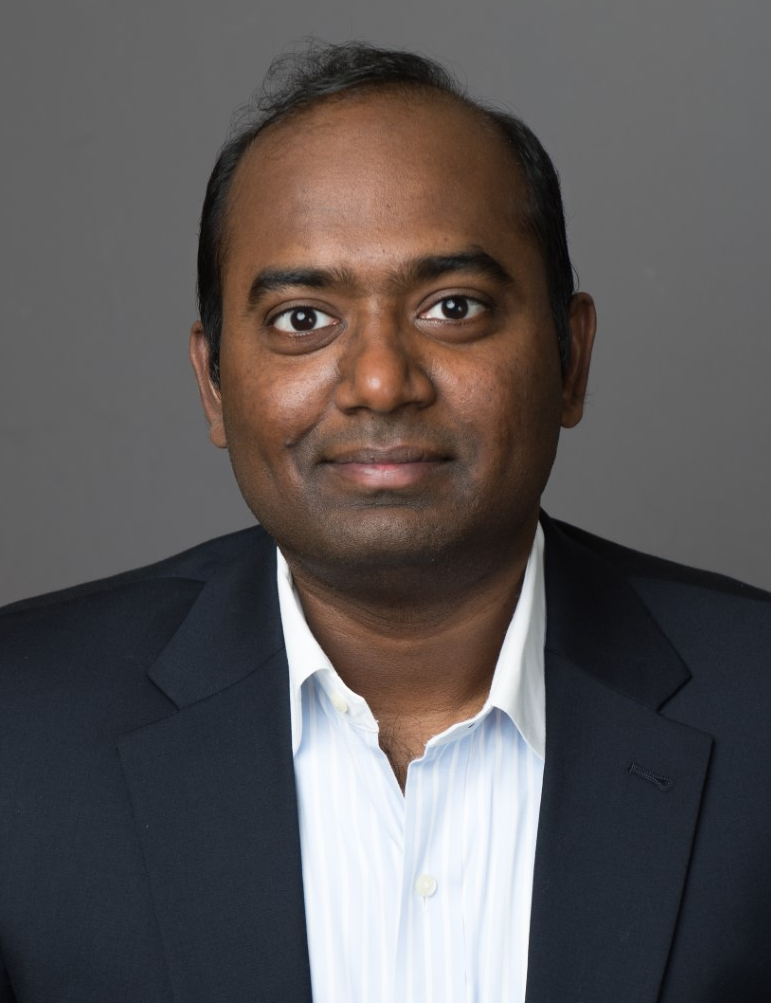}}]{Satish~V.~Ukkusuri}
is a Professor in the Lyles School of Civil Engineering and Director of the Urban Mobility Networks and Intelligence Lab at Purdue University. His research is in the area of interdisciplinary transportation networks with current interests in data driven mobility solutions, disaster management, resilience of interdependent networks, connected and autonomous traffic systems, dynamic traffic networks and smart logistics.  He has published more than 350 peer reviewed journal and conference articles on these topics.
\end{IEEEbiography}

\end{document}